\documentclass[11pt]{article}
\usepackage{jcapmod}

\usepackage{framed}
\usepackage{booktabs}
\usepackage[english]{babel}
\usepackage{amsmath,amssymb,amsbsy,amstext, amsthm, simplewick}
\usepackage{hyperref}
\usepackage{graphicx}
\usepackage{amsfonts}
\usepackage{amssymb}
\usepackage{upgreek}
\usepackage{simplewick}
 \usepackage{exscale,relsize}

\usepackage[margin=1cm,labelfont={sf,bf,scriptsize},textfont={sl,scriptsize}]{caption}

\usepackage{colortbl}
\definecolor{lightgreen}{cmyk}{0.2, 0, 0.2, 0.2}
\definecolor{lightgray}{cmyk}{0.1,0.2,0,0.1}
\definecolor{lightgray2}{cmyk}{0.1,0.1,0,0.1}

\makeatletter
\newlength{\apb@width}
\newcommand{\autoparbox}[2][c]{\settowidth{\apb@width}{#2}\parbox[#1]{\apb@width}{#2}}

\makeatother

\numberwithin{equation}{section}

\def\beq{\begin{equation}}
\def\eeq{\end{equation}}

\def\bea{\begin{eqnarray}}
\def\eea{\end{eqnarray}}

\def\d{{\rm d}}

\newcommand\lsim{\mathrel{\rlap{\lower4pt\hbox{\hskip1pt$\sim$}}
        \raise1pt\hbox{$<$}}}
\newcommand\gsim{\mathrel{\rlap{\lower4pt\hbox{\hskip1pt$\sim$}}
        \raise1pt\hbox{$>$}}}

\def\beq{\begin{equation}}
\def\eeq{\end{equation}}
\def\bea{\begin{eqnarray}}
\def\eea{\end{eqnarray}}

\def\d{{\rm d}}

\def\0{{\bf{0}}}

\def\cO{{\cal O}}

\def\d{{\rm d}}

\def\b{{\boldsymbol{b}}}
\def\k{{\boldsymbol{k}}}
\def\q{{\boldsymbol{q}}}
\def\p{{\boldsymbol{p}}}

\def\x{{\boldsymbol{x}}}
\def\y{{\boldsymbol{y}}}

\def\bz{{\boldsymbol{\zeta}}}
\def\cO{{\cal O}}

\DeclareRobustCommand{\SkipTocEntry}[4]{}

\newcommand{\ket}[1]{| #1 \rangle}
\newcommand{\bra}[1]{\langle #1 |}

\begin{document}

\begin{titlepage}

\setcounter{page}{1} \baselineskip=15.5pt \thispagestyle{empty}

\bigskip\

\vspace{2cm}
\begin{center}
{\fontsize{16}{28}\selectfont  \bf Symmetries and Loops in Inflation}
\end{center}

\vspace{0.2cm}

\begin{center}
{\fontsize{13}{30}\selectfont  Valentin Assassi$^{\bigstar}$, Daniel Baumann$^{\bigstar}$, and Daniel Green$^{\blacklozenge, \spadesuit}$}
\end{center}


\begin{center}

\vskip 8pt
\textsl{$^\bigstar$ D.A.M.T.P., Cambridge University, Cambridge, CB3 0WA, UK}
\vskip 7pt
\textsl{$^ \blacklozenge$
Stanford Institute for Theoretical Physics, Stanford University, Stanford, CA 94306, USA}

\vskip 7pt
\textsl{$^\spadesuit$ Kavli Institute for Particle Astrophysics and Cosmology, Stanford, CA 94025, USA}

\end{center}

\vspace{1.2cm}
\hrule \vspace{0.3cm}
{ \noindent \textbf{Abstract} \\[0.2cm] 
In this paper, we prove that the superhorizon conservation of the curvature perturbation $\zeta$ in single-field inflation  holds as an operator statement. This implies that all $\zeta$-correlators are time independent at all orders in the loop expansion.  Our result follows directly from locality and diffeomorphism invariance of the underlying theory. We also explore the relationship between the conservation of $\zeta$, the single-field consistency relation and the renormalization of composite operators. 
\noindent 
}  
 \vspace{0.3cm}
 \hrule

\vspace{0.6cm}
\end{titlepage}

 \tableofcontents

\newpage

\section{Introduction}

Inflationary perturbations are likely to admit a weakly coupled description.  
In particular, the observed near-Gaussianity of the primordial fluctuations suggests that a treatment in terms of  free fields with computably small corrections is applicable.  From this point of view, higher-order corrections in perturbation theory (i.e.~loop corrections) would appear to be unimportant.  On the other hand, there are examples in field theory in which loop corrections do affect the qualitative behavior of weakly coupled systems.  
For instance, sometimes the tree level contribution to a given quantity vanishes for special reasons (e.g.~gauge invariance forbids the decay of the Higgs boson into photons at tree level).  In these cases, loop corrections are the dominant effect (e.g.~the Higgs decays into photons at one-loop).  Moreover, sometimes the coefficients of loop corrections are (naively) infinite (e.g.~the electron self-energy is divergent).  These cases are more subtle because they require us to understand the meaning of the divergences in order to achieve physical results (e.g.~removing ultraviolet divergences may imply renormalization group flow).

Time-dependent loop corrections to the primordial curvature perturbation $\zeta$ would fall into both categories.  First of all, at tree level $\dot \zeta$ vanishes outside the horizon~\cite{Wands:2000dp, Weinberg:2003sw, Rigopoulos:2003ak, Lyth:2004gb, Langlois:2005qp} and therefore any loop corrections that generate $\dot \zeta \neq 0$ would be the leading effect.  Second of all, the putative corrections are expected to scale as $\log a(t)$~\cite{Weinberg:2006ac}  and therefore diverge as we take the scale factor~$a(t)$ to infinity.  Such infrared divergences would have to be understood before reliable predictions could be made.
  In order to sharpen the understanding of inflationary perturbation theory, one would therefore like to develop results that hold beyond the tree approximation~\cite{Weinberg:2005vy, Weinberg:2006ac, Senatore:2009cf}.  One of the most reliable principles for achieving such results is symmetry.

\vskip 4pt
The universe contains a lot of symmetry. 
On large scales and/or early times, the spacetime is invariant under spatial translations and rotations.
This fixes the metric to be of the Friedmann-Robertson-Walker (FRW) form 
\beq
\d s^2 = - \d t^2 + a^2(t) \, \d \x^2\ , \label{equ:FRW}
\eeq
where $\d\x^2$ represents a maximally symmetric three-space (which we will take to be flat space). 
Small fluctuations around the homogeneous background are close to scale-invariant, suggesting additional symmetry in the action for the fluctuations (e.g.~a global time-translation symmetry).  
The time evolution of the FRW spacetime (\ref{equ:FRW}) can be thought of as a spontaneous breaking of an even larger symmetry group.  
This symmetry breaking can be characterized by introducing the Goldstone mode $\pi$ as a perturbation along the broken symmetry, i.e.~a local shift in time $t+\pi(\x,t)$.
Just as in the classic example of the chiral Lagrangian for pions, the effective action for $\pi$ is highly constrained by the non-linearly realized symmetry~\cite{Creminelli:2006xe}.
This approach has been particularly fruitful for describing inflation~\cite{Cheung:2007st} where the time dependence of the couplings for the Goldstone mode are constrained by additional global symmetries.  For single-clock inflation, the Goldstone mode is directly related to the adiabatic fluctuations that are observed in the cosmic microwave background (CMB).
Transforming to comoving gauge, the field $\pi$ is eaten by the metric.
Adiabatic fluctuations are now represented by the curvature perturbation $\zeta$, defined as the isotropic scalar perturbation to the three-metric,
\beq
g_{ij} = a^2(t) e^{2 \zeta(\x,t)} \delta_{ij}\ .
\eeq
  At late times, $\zeta$ non-linearly realizes conformal symmetries on the spatial slice~\cite{Weinberg:2003sw, Hinterbichler:2012nm, Creminelli:2012ed}.  For instance, under dilatations, $\x \mapsto e^\lambda \x$, the curvature perturbation transforms by a shift, $\zeta \mapsto \zeta + \lambda$. 

\vskip 4pt
At the classical level, the symmetries of $\zeta$ have been used to derive several important theorems about single-field inflation.  For example, Maldacena's consistency relation~\cite{Maldacena:2002vr} uses the dilatation symmetry to show that a long-wavelength mode is unobservable and therefore can't induce observable correlations with short-wavelength modes.  Similarly, Weinberg's proof \cite{Weinberg:2003sw} of the conservation of $\zeta$ on superhorizon scales uses the dilatation symmetry as a method for finding solutions to the classical equations of motion.  One might expect that both statements could be promoted to operator statements in a quantum mechanical theory (where the symmetry can be expressed as a Ward identity~\cite{weinberg2005theV1}).  However, by now there are sufficiently many counterexamples to both results that it is clear that neither statement should follow from symmetry alone.

\vskip 4pt
In this paper, we will prove that $\zeta$ is indeed conserved at all-loop order in single-clock inflation. 
Using nothing more than {\it locality} (which forces commutators to vanish outside of the light-cone), we first show that the operator $\dot{\hat \zeta}$ must satisfy an equation of the form
\beq
\dot{\hat\zeta} = f\big[\hat\zeta \hskip 1pt\big]\ , \label{equ:OpEq-Intro}
\eeq
where $f[\hat\zeta \hskip 1pt]$ is a functional of $\hat \zeta$ and its spatial derivatives.
We then use {\it symmetry}~\cite{Weinberg:2003sw, Hinterbichler:2012nm} to constrain the operators appearing in $f[\hat \zeta]$. Non-derivative operators are forbidden by the dilatation symmetry. The remaining operators can be organized according to their scaling behavior as $a \to \infty$.  (To achieve this, we have to define {\it renormalized composite operators}~\cite{weinberg2005theV2, collins1984renormalization}.) We will find that the leading operators on the right-hand side of eq.~(\ref{equ:OpEq-Intro}) vanish as~$a^{-2}$.  This establishes that, in the limit $a \to \infty$ (or on superhorizon scales), all $\zeta$-correlators are time independent at all orders in the loop expansion.

\vskip 6pt
The outline of the paper is as follows:
In Section~\ref{sec:Symmetries}, we review the symmetries of adiabatic fluctuations in general FRW cosmologies. We show that the curvature perturbation $\zeta$ non-linearly realizes conformal symmetries.
We use these symmetries, in Section~\ref{sec:Proof}, to provide an all-orders proof for the conservation of $\zeta$ on superhorizon scales.
An essential part of the proof is defining a renormalization procedure for composite operators in inflationary spacetimes. We relegate a technical discussion of this subtle issue to Appendix~\ref{sec:Renormalization}.
In Section~\ref{sec:Applications}, we comment on the relationship between our proof for the conservation of $\zeta$ and Maldacena's consistency relation.
We state our conclusions in Section~\ref{sec:Conclusions}.

\section{Symmetries of Adiabatic Fluctuations}
\label{sec:Symmetries}




\subsection{Non-Linearly Realized Symmetries}
\label{sec:NL}

Consider an FRW background with a set of matter fields $\bar\psi_m(t)$.
The time dependence of the background spontaneously breaks time diffeomorphisms.
Just as in particle physics, we can define a Goldstone mode $\pi$ as a perturbation of the fields along the broken symmetry, i.e.~a local shift in time.  
This induces {\it adiabatic} fluctuations
\beq
\delta\psi_m(\x,t) = \bar{\psi}_m\big(t+\pi(\x,t)\big) - \bar\psi_m(t)\ . \label{equ:adiabatic}
\eeq
An effective theory for the Goldstone mode~$\pi$ has been constructed in~\cite{Cheung:2007st, Creminelli:2006xe} (for related work see~\cite{Senatore:2010wk, Senatore:2010jy, Baumann:2011su, Baumann:2011nk, Baumann:2011ws, Behbahani:2012be, Nicolis:2011pv, Behbahani:2011it, Gwyn:2012mw, LopezNacir:2011kk, LopezNacir:2012rm}). It is clear that in the case of purely adiabatic fluctuations, the perturbations in the matter sector can be gauged away by performing a time diffeomorphism
\beq
t\mapsto t-\pi(\x,t)\ .
\eeq
The fluctuations are then in the metric only. These metric fluctuations are described most conveniently in {\it comoving gauge} (also called $\zeta$-gauge), defined as
\beq
\delta \psi_m = 0\  \quad{\rm and}\quad g_{ij}(\x,t) = a^2(t)e^{2\zeta(\x,t)}\delta_{ij}\ , \label{equ:metric}
\eeq
where $\zeta$ is the {\it curvature perturbation}~\cite{Bardeen:1983qw, Salopek:1990jq}. Perturbations in $g_{00}$ and $g_{i0}$ are related to $\zeta$ through the Einstein equations~\cite{Maldacena:2002vr}.
For simplicity, we will drop tensor fluctuations throughout, but re-introducing them doesn't affect our conclusions.  In this gauge, the adiabatic mode is characterized by $\zeta(\x,t)$ directly.  

From the form of (\ref{equ:metric}), we see that the adiabatic mode is invariant under the following {\it large gauge transformations}\hskip 1pt\footnote{By {\it large} gauge transformations we mean gauge transformations that do not vanish at infinity.} \cite{Hinterbichler:2012nm}\hskip 2pt:
\begin{align}
\mbox{dilatation} : \qquad \x &\mapsto \tilde \x \equiv \x e^{\lambda}\ ,\quad \ \zeta(\x) \mapsto \zeta(\tilde \x) + \lambda \ , \label{equ:LG1} \\
\mbox{SCTs} : \qquad \x &\mapsto \tilde \x \equiv \x + 2 (\b \cdot \x) \x - x^2 \b\ ,\quad \
\zeta(\x) \mapsto \zeta(\tilde \x) + 2\hskip 1pt \b \cdot \x \ , \label{equ:LG2}
\end{align}
where SCT stands for special conformal transformation. Notice that $\zeta$ transforms non-linearly: dilatations shift the value of $\zeta$, while SCTs shift its spatial gradient.
Both of these symmetries are part of the group of diffeomorphisms under which the theory is invariant.  What makes the transformations in (\ref{equ:LG1}) and (\ref{equ:LG2}) special is the fact that they preserve $\zeta$-gauge, but are not removed by gauge fixing. 
After gauge fixing, the large gauge transformations therefore remain a symmetry of the action.  As for any global symmetry, this implies the presence of  conserved currents: one for the dilatation, $J_{d}^\mu$, and three  for the special conformal transformations, $J_{sc\,(i)}^\mu$. 
In the following, we will drop the subscripts whenever an expression applies to both types of currents and keep it only when a distinction needs to be made. Current conservation, $\partial_\mu J^\mu = 0$, implies the following Ward identity~\cite{weinberg2005theV1} for correlation functions~\cite{Assassi:2012zq}
 \begin{align}
& i\, \partial_\mu^{(x)} \big\langle J^{\mu}(\x,t) \zeta(\y_1, t_\star) \cdots \zeta(\y_n, t_\star) \big\rangle \ = \ \nonumber \\
&\hspace{0.5cm}\ = \ \sum_{i=1}^n \delta(t-t_\star)\delta({\x}-{\y}_i \hskip 1pt ) \hskip 2pt \big\langle \zeta(\y_1, t_\star) \cdots \delta \zeta({\y_i}, t_\star) \cdots \, \zeta({\y}_n, t_\star) \big\rangle  \ , \label{equ:Ward0}
\end{align}
where $\delta \zeta$ denotes infinitesimal variations of $\zeta$ under the large gauge transformations
\bea
\delta_d\hskip 1pt \zeta &\equiv&  -1- \x \cdot \partial_\x \hskip 1pt  \zeta\ , \\
\delta^{(i)}_{sc}\hskip 1pt \zeta &\equiv&   - 2 x^i - 2 x^i (\x\cdot \partial_\x \hskip 1pt  \zeta)  + x^2 \partial^i \hskip 1pt  \zeta \ .
\eea
Here, we have introduced an index $i$ to distinguish the three SCTs associated with the three components of the vector $\b$.  
Finally, it is also convenient to define a conserved charge associated with each symmetry
\beq
Q = \int \d^3 x \, J^0 \ .
\eeq
Formally, this satisfies $\dot Q = 0$.  However, when the symmetry is spontaneously broken, IR divergences make the value of $Q$ ill-defined. On the other hand, $Q$ remains well-defined in commutators with local operators, such as $[Q, \zeta]$, and inside correlation functions.  In fact, by integrating the Ward identity (\ref{equ:Ward0}) for $n=1$, we see that 
\beq\label{equ:chargecomm}
i [Q, \zeta ] = \delta \zeta \ .
\eeq 

\subsection{Symmetries and the Conservation of Zeta}
\label{sec:Zeta}


The presence of the dilatation symmetry has played a crucial role in previous work on the constancy of $\zeta$ outside the horizon.  At a technical level, this connection was implemented most directly by Weinberg~\cite{Weinberg:2003sw}, who used the existence of the large gauge transformation to find two physical solutions to the classical equations of motion in any FRW background: one solution is a constant and the other decays as $a^{-3}$. These two solutions correspond to the growing and decaying contributions of the adiabatic mode.  If we assume that only the adiabatic mode is present, 
then we have found all the possible solutions and therefore $\zeta$ is conserved classically.

\vskip 4pt
\noindent
{\it Tree-level.}---Although Weinberg used Newtonian gauge, his result is easily reproduced from the dilatation symmetry in $\zeta$-gauge.  For our purposes, it will be useful to state Weinberg's proof in a quantum mechanical language using the Ward identity (\ref{equ:chargecomm}).
Taking the expectation value, we find
\beq\label{equ:classicalmode1}
\big\langle \big[Q_{d}, \zeta_{\k} \big] \big\rangle = i (2\pi)^3\delta(\k)\ .
\eeq
Since $\dot Q_d = 0$, the time derivative of this expression is 
\beq\label{equ:classicalmode2}
\big\langle \big[Q_d ,\dot \zeta_{\k} \big] \big\rangle = 0\ .
\eeq  
In order to satisfy (\ref{equ:classicalmode1}), we require a non-zero solution for $\zeta_{\k \to 0}$, while (\ref{equ:classicalmode2}) implies that this solution is time independent.  We have therefore found that a non-zero constant is a solution for $\zeta_{\k \to 0}$.  
Moreover, locality requires that (see \S\ref{sec:locality})
\beq
\big[\dot \zeta_\k(t), \zeta_{\k'}(t) \big] \propto a^{-3}(t) \hskip 1pt (2\pi)^3\delta(\k+\k')\ .
\eeq 
This implies the existence of a second solution scaling as $a^{-3}$.  Since there are only two solutions to the classical equations of motion, we have found that $\zeta$ is classically conserved.\footnote{Technically speaking, we have not shown that these solutions can be extended to finite momentum $k$.  However, using the Ward identity (\ref{equ:Ward0}) it is straightforward to prove that this is the case (see Appendix~A of \cite{Assassi:2012zq}).} 
In this paper, we will extend Weinberg's proof to the quantum level.

\vskip 4pt
\noindent
{\it One-loop.}---It is well-known that massless scalar fields can receive time evolution outside the horizon from quantum corrections.
Essentially, this arises because radiative corrections induce a mass for any unprotected scalars, which then sources superhorizon evolution~\cite{Senatore:2009cf}. Two-point functions are found to evolve as $\log a(t)$. It is therefore natural to ask what happens  to the conservation of $\zeta$ at loop level. This question was first raised by Weinberg in \cite{Weinberg:2005vy, Weinberg:2006ac}.
Subsequently, a calculation by Kahya, Onemli and Woodard~\cite{Kahya:2010xh}, indeed, suggested that loops would induce a time dependence of~$\zeta$.
This conclusion was challenged by Pimentel, Senatore and Zaldarriaga~\cite{Pimentel:2012tw}.
In an impressively complex calculation, these authors showed that although individual one-loop diagrams do induce a time dependence, the effect precisely cancels when {\it all} diagrams are summed.
Not surprisingly, symmetry played an important role in understanding this cancellation.  At various stages in their calculation Pimentel, Senatore and Zaldarriaga, directly or indirectly, employed the dilatation symmetry.
They showed that a class of diagrams sums to zero on account of the single-field consistency relation~\cite{Maldacena:2002vr,Creminelli:2004yq} (which is closely related to the Ward identity in (\ref{equ:Ward0}); see~\cite{Assassi:2012zq} and Section~\ref{sec:Applications}), while others cancel because they are related by the non-linear transformation of~$\zeta$.  

These types of cancellations are reminiscent  of those appearing in QED. For instance, consider photon-photon scattering. The leading-order diagram contains four external photons and a fermion loop connecting them. Each individual diagram, corresponding to a particular permutation of legs, is logarithmically divergent. However,  the divergences exactly cancel when all diagrams are summed.  In this case, the cancellation is, of course, a consequence of gauge invariance. To see this, consider the amplitude ${\cal M}_{\mu\nu\sigma\rho}$,
which by Lorentz invariance takes the following form
\beq
{\cal M}_{\mu\nu\sigma\rho} = K(\eta_{\mu\nu}\eta_{\sigma\rho}+\eta_{\mu\sigma}\eta_{\nu\rho}+\eta_{\mu\rho}\eta_{\nu\sigma}) + {\rm finite\ terms}\ .
\eeq
A priori, the amplitude $K$ could be divergent, but the Ward identity, $p^\mu{\cal M}_{\mu\nu\sigma\rho} = 0$, forces it to be finite. This is an important result, since a divergence would have forced us to introduce a $(A_\mu A^{\mu})^2$ counterterm, and consequently break gauge invariance~\cite{weinberg2005theV1}. 
In this paper, we will show that symmetries similarly protect correlation functions of curvature perturbations from getting a late-time evolution.


\vskip 4pt
\noindent
{\it Towards all orders.}---At a qualitative level, it is easy to convince oneself that the dilatation symmetry implies constancy of $\zeta$ to all orders in perturbation theory.  We will ultimately agree with this intuition (see Section~\ref{sec:Proof}), but we would first like to point out where we feel that some details are missing.  This may explain why some authors have not been convinced by these arguments.

A general sentiment one encounters in the literature is that, because a constant $\zeta$ mode can be removed by a gauge transformation, $\zeta_{\k \to 0}$ cannot be the source for a time-dependent solution. 
However, this argument appears somewhat circular since a time-dependent mode $\zeta(t)$ cannot be removed by such a transformation.
On the other hand, one might have imagined that a time dependence of $\zeta$ would require an operator equation of the form $\dot \zeta = c_1 \zeta+ c_2 \zeta^2 + \cdots$.  The right-hand side of this equation is incompatible with the dilatation symmetry and is therefore forbidden to act as a source for $\dot \zeta$.  However, why should such an operator equation be the only possibility?
Moreover, a trivial counterexample to this logic is the case of a Goldstone boson, $\pi$, which transforms as $\pi \mapsto \pi + 1$.  As in the case of $\zeta$, a constant value of $\pi(\x,t) = \pi_0$ is unphysical because it can be removed by a global transformation $\pi \mapsto \pi - \pi_0$, which simply moves us between equivalent vacua.  However, in flat space, the conclusion that $\dot \pi_{\k \to 0} = 0$ as an operator is clearly false because quantum mechanics requires that $[\dot \pi_{\k} , \pi_{\k'} ] = i (2\pi)^3\delta(\k+\k')$.  
Of course, this counterexample isn't quite a fair analogy since in the case of $\zeta$ we know that the modes become classical outside the horizon and freeze at tree level.
Nevertheless, the example does illustrate that the argument has to involve more than symmetry alone.

A more serious concern is that modes inside the horizon could induce a coherent effect on large scales that would cause a time dependence of the long-wavelength modes~\cite{Weinberg:2005vy, Weinberg:2006ac}. These short-scale modes are physical and cannot be removed by any symmetry. Hence, such coherent effects cannot be argued to vanish by symmetry alone.  For example, time dependence could, in principle,  arise from $\dot \zeta(\x,t) = c\, \partial_i \zeta \hskip 1pt \partial^i \zeta(\x,t)$, which is compatible with the dilatation symmetry.  In momentum space, this becomes 
\beq
\dot \zeta_\k = c \int \frac{\d^3 p}{(2\pi)^3} \, \frac{\p \cdot (\k -\p)}{a^2} \hskip 1pt \zeta_\p\hskip 1pt \zeta_{\k-\p}\ , \label{equ:GradSource}
\eeq 
which receives contributions from $p \gtrsim a H$.  The source term in eq.~(\ref{equ:GradSource}) is not suppressed as $a \to \infty$ since the momentum integral formally includes contributions from $p \to \infty$.  Of course, the treatment of these effects is complicated by the fact that the integral is UV divergent and needs to be regulated. One has to be careful that a bad choice of regulator doesn't introduce a spurious time dependence for~$\zeta$.

\vskip 6pt
Having described some of the subtleties involved in the conservation of $\zeta$ at loop level, we will, in the next section, provide an all-orders symmetry-based proof for the time-independence of $\zeta$-correlators.

\section{A Non-Renormalization Theorem} 
\label{sec:Proof}

Our proof involves just a few relatively straightforward steps.
First, we will prove that the modes always become classical outside the horizon (\S\ref{sec:locality}; see also \cite{Lyth:2006qz}).  We will show that this implies that any time evolution outside the horizon is described by an operator equation of the form\footnote{To avoid confusion, we will (in this section only) use a hat to
denote quantum operators and reserve unhatted variables for c-numbers (such as the eigenvalues of $\hat \zeta$).}~(\S\ref{sec:evolution})
\beq
\dot{\hat \zeta}(\x,t) = \sum_n \alpha_n(t) \hat \zeta^n (\x,t)+ \cdots \ . \label{equ:OPEQ}
\eeq
We will then use the dilatation symmetry to show that $\alpha_n = 0$ (\S\ref{sec:symmconstraints}).
   Finally, we will show that the additional terms ($\cdots$) vanish at least like powers of $a^{-2}$ and therefore can be ignored at late times. 
  To understand this power law suppression requires a careful treatment of the renormalization of composite operators (see \S\ref{sec:subcomposite} and Appendix~\ref{sec:Renormalization}).
   These terms include the effects of the modes inside the horizon that had been the concern of previous authors~\cite{Weinberg:2005vy, Weinberg:2006ac}.  
   
   Any no-go result is only as good as its assumptions. Let us therefore be clear about the assumptions that go into our proof: 
   First, we will assume throughout that the theory is local and that the initial state is the Bunch-Davies vacuum. 
   Second, we will only address loop corrections during inflation, such that the mode functions for the interaction pictures fields are roughly the de Sitter solutions.  Finally, we will assume that any time-dependent couplings in the action for $\zeta$ scale at most like $(\log a(t))^r$, for some finite $r$, and not as powers of $a(t)$. 
   The last two assumptions are mostly of technical nature and can probably be relaxed.
However, even with these simplifying assumptions, our analysis is sufficiently general to cover the vast majority of inflationary models. We comment on ways to circumvent our theorem in \S\ref{sec:results}.

\subsection{Locality and Classicality}
\label{sec:locality}

We begin by establishing the relation between locality of the theory and classicality of $\zeta$ on superhorizon scales.
We will define a mode $\zeta_\k$ as being ``classical" at late times, if it satisfies
\beq
{\cal C} \equiv 
\frac{\big\langle \big[\dot{\hat \zeta}, {\hat \zeta}\big]\big\rangle^2}{\big\langle \dot{\hat \zeta}^2 \big\rangle \big\langle {\hat \zeta}^2 \big\rangle}\ \xrightarrow{a\to \infty}\ 0\ , \label{equ:classical}
\eeq
where all the operators are evaluated at the same time.  This definition of classicality implies that equal-time correlation functions of $\hat \zeta_\k$ and/or $\dot{\hat\zeta}_\k$ can be rewritten in terms of classical stochastic variables, up to corrections that vanish as $a \to \infty$---i.e.~we can ignore all commutators at sufficiently late times.  Moreover, eq.~(\ref{equ:classical}) assumes that the theory is approximately Gaussian, so that the power spectrum can be used to estimate of the size of any correlation function.

Locality severely constrains the possible forms of equal-time commutators, like the one that appears in (\ref{equ:classical}). 
In particular, the commutator of any pair of local operators $\hat \cO_1$ and $\hat \cO_2$ must satisfy 
\beq
\big[\hat \cO_1(\x,t), \hat \cO_2(\y,t) \big] = 0\ , \qquad {\rm for} \quad \x \neq \y\ .
\eeq
As a result, the commutator must be proportional $(\sqrt{-g} \hskip 1pt)^{-1}\hskip 1pt \delta(\x-\y)$ or derivatives therefore, 
\beq\label{equ:commutator}
\big[\dot{\hat \zeta}(\x,t), {\hat \zeta}(\y,t)\big] = \left[\sum_n c_n(t) {\hat \cO}^{(n)}(\x,t) + \Big(\sum_m d_m(t) {\hat \cO}^{(m)}_i (\x,t)\Big) g^{ij} \partial_j  + \cdots  \right] \frac{\delta(\x-\y)}{\sqrt{-g}}   \ ,
\eeq
where ${\hat \cO}^{(n)}$ and ${\hat \cO}^{(m)}_i$ are some basis of local scalar and vector operators, respectively. 
If the action for $\zeta$ is time independent, then the coefficients in (\ref{equ:commutator}) must be time independent as well, i.e.~$c_n(t) \to c_n$ and $d_m(t) \to d_m$.  Similarly, if the couplings in the action scale like $(\log a(t))^r$ for some finite $r$, then the coefficients in (\ref{equ:commutator}) are also logarithmic in $a(t)$.    In the limit $a\to \infty$, we see that (\ref{equ:commutator}) therefore vanishes at least as $a^{-3}$, due to the overall factor of $(\sqrt{-g}\hskip 1pt)^{-1} = a^{-3}$ required by diffeomorphism invariance.

To establish that the mode becomes classical in the sense of eq.~(\ref{equ:classical}), we now show that $\langle \zeta^2 \rangle$ is bounded from below by a constant as $a \to \infty$.  First, let us insert a complete set of states into (\ref{equ:classicalmode1}), 
\beq\label{equ:constantmode1}
\langle Q_d \hskip 1pt \zeta_\k \rangle = \sum_n \langle Q_d | n \rangle \langle n | \zeta_\k \rangle  = \langle Q_d | 1 \rangle \langle 1 | \zeta_\k \rangle = \frac{i}{2} (2\pi)^3 \delta(\k) \ ,
\eeq
where we have rotated the basis of states such that $\langle Q_d | n \rangle = \delta_{n 1} \langle Q_d | 1\rangle$.  Similarly, we can insert the same set of states into the power spectrum of $\zeta$ to find
\beq\label{equ:constantmode2}
\langle \zeta_\k \zeta_{-\k} \rangle = \sum_n | \langle \zeta_\k | n \rangle|^2 \, \geq\, |\langle\zeta_\k |1\rangle|^2 \ .
\eeq
Using $\dot Q_d = 0$ and assuming\footnote{This is essentially the assumption that a generalization of the Goldstone boson decay constant, $f_\pi$, associated with $Q_d$ is finite.  This is equivalent to demanding that $\zeta$ is dynamical, i.e.~has a finite kinetic term.} $|\langle Q_d | 1 \rangle |< \infty$, we must have $ | \langle \zeta |1\rangle|^2 > \xi > 0$ where $\xi$ is a constant.  We see that ${\cal C} \to 0$ as $a \to \infty$ provided that $\dot \zeta$ vanishes more slowly than $a^{-3}$.
Recall that the goal of this section is to prove that $\dot \zeta$ vanishes at least as $a^{-2}$. Anything that violates our definition of classicality vanishes even faster.
In that case, there is nothing for us to prove.

\subsection{Operator Evolution}
\label{sec:evolution}

We have proven that the modes of interest become classical at late times.  If these were solutions to the classical equations of motion for a single degree of freedom, then they would be determined by two boundary conditions. 
For the problem at hand, one boundary condition is set by the choice of the Bunch-Davies vacuum and the other can be chosen to be the classical field configuration for $\zeta(\x,t)$ at a later time $t$.  Therefore, given $\zeta(\x,t)$, the classical soultion for $\dot \zeta(\x,t)$ is fixed.  The purpose of this subsection is to make this statement precise, as an operator equation 
\beq
\dot{\hat \zeta}(\x,t) = f(\hat \zeta(\x,t), \partial \hat \zeta(\x,t), \cdots)\ ,  \label{equ:OpEq}
\eeq
for some functional $f[\hat \zeta]$.

\vskip 6pt
\noindent
{\it Simultaneous eigenstates.}---As in any quantum field theory describing a single degree of freedom, the operators $\hat \zeta(\x,t)$ form, at any time $t$, a complete set of commuting observables (one for each point in space). This has two important consequences~\cite{DiracBook}:
\begin{itemize}
\item First, the eigenstates of these operators, $\ket{\zeta(\x,t)}$, are non-degenerate and form a complete basis of states on the Hilbert space.\footnote{The relation between these states and the classical solutions is most transparent in the Schr\"odinger picture, where we define a wavefunction for $\zeta(\x,t)$, i.e.~$\Psi[\zeta(\x,t)]$.  For a single degree of freedom, the wavefunction satisfies a differential equation whose solution is determined by the initial state in the far past (e.g.~Bunch-Davies) and the field configuration $\zeta(\x, t)$ at late times $t$.  See \cite{Maldacena:2002vr} for more details on the connection between $in$-$in$ calculations and the Schr\"odinger representation.}
\item Second, any operator $\hat \cO(\y,t)$ which commutes with  $\hat \zeta(\x,t)$ is a function of $\hat\zeta(\x,t)$ alone, i.e.~we have $\hat \cO(\y,t) = f[\hat \zeta(\y,t)]$, where $f$ is a functional of $\hat \zeta$.
\end{itemize}
Since at late times the commutator of $\dot{\hat\zeta}$ and $\hat\zeta$ vanishes (see eq.~(\ref{equ:classical})), we expect that, in the limit $a(t)\rightarrow\infty$, the operator $\dot{\hat\zeta}$ can be written as a function of $\hat\zeta$. Let us derive this result more formally.
We start by defining the basis of eigenstates of $\hat\zeta$ as 
\beq
\ket{\bz} \equiv |(\zeta_1, \zeta_2, \cdots, \zeta_a, \cdots) \rangle\ , \qquad {\rm where} \quad \zeta_a \equiv \zeta(\x_a,t)\ . \label{equ:ZetaBasis}
\eeq
For clarity, we have used a discrete index to denote the spatial position. 
The state $\ket{\bz}$ is, by definition,  an eigenstate of the operator $\hat \zeta_a$ with eigenvalue $\zeta_a$, i.e.
\beq
\hat \zeta_a \ket{\bz } = \zeta_a \ket{\bz}\ .
\eeq	
In this notation, the commutator (\ref{equ:commutator}) becomes
\beq
\big[\dot {\hat \zeta}_a , \hat \zeta_b \big] = \frac{\hat A_a}{a^3(t)} \delta_{ab} + \cdots\ . \label{equ:COM}
\eeq
where $\hat A_a \equiv \sum_n c_n(t) \hat \cO^{(n)}_a(t)$ and the ellipses denote terms that are suppressed by additional powers of $a(t)$.  Evaluating eq.~(\ref{equ:COM}) in the $\zeta$-basis (\ref{equ:ZetaBasis}), we find 
\beq
(\zeta_b - \tilde \zeta_b) \big \langle \tilde \bz \big| \dot {\hat \zeta}_a \big | \bz\big \rangle = \frac{\big \langle \tilde \bz \big | \hat A_a \big | \bz \big \rangle}{a^3(t)} \delta_{ab} + \cdots\ . 
\eeq
This equation defines $ \dot {\hat \zeta}$ as an operator since it allows us to compute any matrix element by inserting a complete set of states.  The r.h.s.~of this equation scales at least as $a^{-3}$ (up to $\log a$ corrections) and therefore the leading behavior is governed by the homogeneous solution, namely 
\beq\label{equ:homogeneous}
\big \langle \tilde \bz \big| \dot {\hat \zeta}_a \big | \bz \big\rangle \, \approx\,  f_a[\bz] \hskip 1pt  \delta(\tilde \bz - \bz) + {\cal O}(a^{-3}) \ ,
\eeq
where $\delta(\tilde \bz - \bz) \equiv \prod_a \delta(\tilde \zeta_a - \zeta_a)$ and ${\cal O}(a^{-3})$ stands for operators whose correlation functions vanish as $a \to \infty$ (we make this more precise in Appendix~\ref{sec:Renormalization}).

\vskip 6pt
\noindent
{\it Locality.}---Next,  let us see how locality constrains the form of the functional $f_a[\bz]$. 
Recall that the conjugate momentum $\hat \Pi$ satisfies the canonical commutation relation $[\hat \zeta_a,\hat \Pi_b ] = i \delta_{ab}$.  Locality also requires that $[\dot {\hat \zeta}_a,\hat \Pi_b] \propto \delta_{ab}$.  Together with eq.~(\ref{equ:homogeneous}), we then find 
\beq
\big\langle \tilde \bz\big| \big[\dot {\hat \zeta}_a,\hat \Pi_b\big]  \big| \bz \big\rangle = -i  \frac{\partial f_a[\bz]}{\partial {\zeta_b}}  \delta(\tilde \bz - \bz) \propto \delta_{ab}\ .
\eeq
As a result, $f_a[\bz]$ cannot depend explicitly on $\zeta_b$ for $b \neq a$.  Invariance under spatial translations furthermore implies that $f_a[\zeta_a] = f[\zeta_a]$. 
We have therefore established that
\beq\label{equ:funczeta}
\bra{\tilde \bz} \dot {\hat \zeta}_a \ket{\bz} \approx f(\zeta_a, \partial^i \zeta_a, \cdots)\hskip 1pt \delta(\tilde \zeta_a - \zeta_a)\ .
\eeq
Although this is a statement involving matrix elements in the $\zeta$-basis, the result holds in any basis.  To see this, note that (\ref{equ:funczeta}) holds inside any correlation function:
\bea
\langle \dot {\hat \zeta}_a \ \cdots \rangle &=& \int \d \tilde \bz \int \d \bz\,  \bra{0} \tilde \bz \rangle \bra{\tilde \bz} \dot {\hat \zeta}_a \ket{\bz} \bra{\bz}\, \cdots \rangle = \langle f[\zeta_a] \, \cdots \rangle \ .
\eea
This proves that (\ref{equ:funczeta}) is equivalent to the operator statement
\beq
\dot {\hat \zeta}(\x,t) = f[\hat \zeta(\x,t)]  \label{equ:func}  \ ,
\eeq
where we have dropped the terms of order $a^{-3}$ coming from the non-zero commutator (\ref{equ:COM}). 
We will study the implications of this equation in the next two subsections.

\subsection{Constraints from Symmetry}\label{sec:symmconstraints}

 A basic property of any operator equation is that the two sides of the equation must transform in the same way under symmetries.  In this subsection, we will show that the symmetries of $\zeta$ (see Section~\ref{sec:Symmetries}) severely constrain which operators are allow to appear on the r.h.s.~of eq.~(\ref{equ:func}). 

\vskip 4pt
\noindent
{\it Perturbation expansion.}---In perturbation theory, we usually consider situations where $\langle \zeta^2 \rangle \ll 1$.  This corresponds to the requirement that the split of the metric into background and fluctuations is reliable.  Given the small amplitude of fluctuations and the assumption of weak coupling, 
we can Taylor expand the r.h.s.~of eq.~(\ref{equ:func}) around $\zeta = 0$,
\beq\label{equ:taylor}
\dot {\hat \zeta} (\x,t) = \sum_{n=0}^{\infty} \alpha_n(t) \hat \zeta^n(\x,t) + \sum_{m=1}^\infty \beta_m(t) \big(a^{-2} e^{-2 \hat \zeta}\partial^2 \hat \zeta(\x,t) \big)^m + \sum_{\ell=1}^\infty \gamma_\ell(t) \big(g^{ij} \partial_i \hat \zeta \partial_j \hat\zeta (\x,t) \big)^{\ell} + \cdots \ .
\eeq

\vskip 4pt
\noindent
{\it Renormalization condition.}---Since $\hat\zeta$ is a fluctuation, we require the expectation value of the l.h.s.~of (\ref{equ:taylor}) to vanish, $\langle\hskip 1pt \dot {\hat \zeta}  \hskip 2pt\rangle  = 0$.  This fixes the coefficient of the unit operator, $\alpha_0(t)$, in terms of the vacuum expectation values of the other local operators.  Of course, we are always free to define $\langle \cO \rangle \equiv 0$ as a renormalization condition for all local operators $\cO \neq \hat 1$.  
In that case, the coefficient of the unit operator must vanish, $\alpha_0(t) = 0$.

\vskip 4pt
\noindent
{\it Dilatation symmetry.}---Next, we consider the constraints imposed by the dilatation symmetry. 
Recall that $i [\hat Q_d, \hat \zeta] = -1 - \x \cdot \partial_\x \hat \zeta$, which means that 
\beq
i \big[\hat Q_d , \dot {\hat \zeta} \big] = - \x \cdot \partial_\x \dot {\hat \zeta}  \ .
\eeq
The higher-derivative operators $\hat \cO^{(\partial)}$ in eq.~(\ref{equ:taylor}) (i.e.~those with coefficients $\beta_m$, $\gamma_\ell$, etc.) have been arranged in such a way that $i [\hat Q_d, \hat \cO^{(\partial)}] =- \x \cdot \partial_\x \hat \cO^{(\partial)}$.  Therefore, any values of the coefficients $\beta_m$ and $\gamma_\ell$ are consistent with the transformation of $\dot {\hat \zeta} (\x,t)$ under $\hat Q_d$.  The same is not true for the operators $\hat \zeta^n$, which transform as 
\beq
i \big[\hat Q_d, \hat\zeta^n \big] = -n \hat \zeta^{n-1} - \x \cdot \partial_\x \hat \zeta^n \ \neq\  - \x \cdot \partial_\x \hat \zeta^n\ .
\eeq  
We see that each individual term in the sum over $\hat \zeta^n$ does not transform correctly to match the transformation of $\dot {\hat \zeta}$.  Furthermore, there is no way to choose the coefficients $\alpha_n$ in such a way that the additional terms in the transformations of $\hat \zeta^n$ cancel between terms.  
Therefore, consistency with the transformation under $\hat Q_d$ requires that $\alpha_n =0 $ for all $n$.

\vskip 4pt
\noindent
{\it Special conformal symmetry.}---We can repeat the same analysis for the SCTs generated by $\hat Q_{sc}^i$.  From $i [\hat Q_{sc}^i, \hat \zeta] =  - 2 x^i - 2 x^i (\x\cdot \partial_\x \hskip 1pt  \hat \zeta \hskip 1pt  )  + x^2 \partial^i \hskip 1pt \hat \zeta$, we infer that
\beq
i \big[\hat Q_{sc}^i,\dot {\hat \zeta} \big] =  - 2 x^i \big(\x\cdot \partial_\x \hskip 1pt  \dot {\hat \zeta} \hskip 1pt  \big)  + x^2 \partial^i \hskip 1pt  \dot {\hat \zeta} \ .
\eeq
Matching the transformation on the r.h.s.~of (\ref{equ:taylor}) imposes non-trivial relations between the coefficients.  For example, at second order in derivatives we have
\beq
\dot {\hat \zeta} (\x,t) =  a^{-2} e^{-2 \hat \zeta} \big(\hskip 1pt \beta_1(t) \partial^2 \hat \zeta + \gamma_1 (t) \delta^{ij} \partial_i \hat \zeta \partial_j \hat \zeta  \hskip 1pt \big) + \cO(\partial^4) \ .
\eeq
Imposing that the transformations on both sides agree gives 
\beq
\dot {\hat \zeta} (\x,t) =  \beta_1(t)  \hskip 1pt a^{-2} e^{-2 \hat \zeta}\big( \hskip1pt \partial^2 \hat \zeta + \tfrac{1}{2}  \delta^{ij} \partial_i \hat \zeta \partial_j \hat \zeta \hskip 1pt \big) + \cO(\partial^4) \ .
\eeq
The special combination of operators on the r.h.s.~should not be too surprising, since it is precisely the combination that appears in the three-dimensional Ricci scalar, 
\beq
{\cal R} \equiv - 4 \hskip 1pt a^{-2} e^{-2 \zeta}\big( \hskip1pt \partial^2  \zeta + \tfrac{1}{2}  \delta^{ij} \partial_i \zeta  \partial_j  \zeta \hskip 1pt \big)\ .
\eeq  
Both the dilatation and the SCTs are continuously connected to a general, time-independent diffeomorphism on the spatial slice.  Because $\dot {\hat \zeta}$ transforms as a scalar under this group, the r.h.s.~of (\ref{equ:taylor}) should be composed of invariants of the group.  For this reason, our equation should take the form 
\bea\label{equ:taylorcurvature}
\dot {\hat \zeta} (\x,t) 
&=& \tilde \beta_1(t)\hskip 1pt  \hat{\cal R} (\x,t)+ \tilde \beta_2(t)\hskip 1pt \hat{\cal R}^2(\x,t) + \tilde \beta_3(t)\hskip 1pt \hat{\cal R}_{ij} \hat{\cal R}^{ij}(\x,t)+\cdots \ ,
\eea
where ${\cal R}_{ij}$ is the Ricci tensor on the spatial slice.
The final step in our proof will be to show that all these terms vanish at least like powers of $a^{-2}$.  Phrased in terms of curvatures, it seems  intuitive that inflation should smooth out the spatial curvatures.  Showing that this intuition survives quantum corrections will be the subject of the next subsection.

\subsection{Renormalization of Composite Operators}\label{sec:subcomposite}

The right-hand-side of eq.~(\ref{equ:taylorcurvature}) contains composite
operators, i.e.~products of fields evaluated at coincident points, which even in a Gaussian theory leads to divergences. One might worry that these divergences will affect the scaling behaviour of the operators at late times, i.e.~change their $a^{-n}$ suppression. In particular, the renormalization of composite operators is complicated by the tendency of operators to mix under
renormalization~\cite{collins1984renormalization, weinberg2005theV2}. 
In order to complete the proof, we need to show that the scaling of the operators in (\ref{equ:taylorcurvature}) isn't  drastically affected by renormalization.
More precisely, we wish to show
 that if a composite operator $\cO(\x,t)$ scales like $a^{-n}$ in the free theory, any corrections in the interacting theory that scale like $a^{-m}$, where $m$ is an integer with $m < n$, can be removed by a local redefinition of the operator, 
 \beq
 \cO_R(\x,t) \equiv \cO(\x,t) + \delta \cO(\x,t)\ .
 \eeq
This allows us to define renormalized composite operators by local subtraction.
By definition, these operators then all decay at least as powers of $a^{-2}$ in correlation functions.  

\vskip 4pt
An explicit demonstration of the renormalization of composite operators by local subtraction is rather technical. 
In this subsection, we therefore only show how the renormalization works in a concrete example (see also \cite{Harlow:2011ke} for a related discussion). The dedicated reader can find the painful details for the most general cases in Appendix~\ref{sec:Renormalization}.

\subsubsection*{Example}
\vskip 4pt
Consider a massless scalar field\footnote{Since the renormalization of composite operators is unrelated to the special symmetries satisfied by $\zeta$, we have switched to a generic scalar field $\phi$. } in de Sitter space with interaction $\dot \phi^3$.  
For purposes of illustration, we will present the renormalization of the composite operator\footnote{For notational simplicity, we will sometimes drop the time argument, i.e.~${\cO}(\x)$ means $\cO(\x,t)$.} $\cO(\x) = (\partial^2 \phi/a^2)^2(\x)$.  

\vskip 4pt
\noindent
{\it Tree-level scaling.}---Even in the free (or Gaussian) theory, this operator has a non-vanishing one-point function 
\beq
\langle \cO \rangle =  \int^{a \Lambda} \frac{\d^3 k}{(2\pi)^3} \frac{k^4}{a^4} |\phi_\k|^2 = \frac{H^2}{4\pi^2} \int^{a\Lambda}  \frac{k^3 \d k}{a^4}\left( 1+ \frac{k^2}{(aH)^2}\right) = \frac{\Lambda^4}{16 \pi^2}\left[  H^2 + \frac{2}{3}\Lambda^2 \right] \ .
\eeq
We have cut off the integral at fixed physical momentum $\Lambda$ and used the Bunch-Davies mode function
\beq
\phi_\k(\tau) = \frac{H}{\sqrt{2k^3}} (1+ ik\tau) e^{-ik\tau}   \ , 
\eeq
where $\tau$ is conformal time.
In accordance with our renormalization condition, we define a shifted operator with vanishing one-point function,
\beq
\cO_R \equiv \cO - \langle \cO \rangle\ .
\eeq
Next, let us consider the two-point function of this operator (still in the Gaussian theory)
\bea
\langle   \cO_R(\x)  \cO_R(\0) \rangle 
&=&   \frac{2}{a^8} \big( \big\langle \partial^2 \phi(\x) \partial^2 \phi(\0) \big\rangle \big)^2 \ , 
\eea
or 
\begin{align}
\int \d^3 x \, e^{i \k \cdot \x} \, \langle   \cO_R(\x)  \cO_R(\0) \rangle  &=  \frac{2}{a^8}  \int^{a\Lambda} \frac{\d^3 q}{(2\pi)^3} \, q^4 |\phi_\q|^2  |\k-\q\hskip 1pt|^4 |\phi_{\k-\q}|^2\nonumber \\
&=
\frac{H^4}{2 a^8} \int^{a\Lambda} \frac{\d^3 q}{(2\pi)^3}\, q |\k - \q\hskip 1pt| \left(1+ \frac{q^2}{(aH)^2} \right)  \left(1+ \frac{|\k-\q\hskip 1pt|^2}{(aH)^2} \right)  \nonumber \\
&=-  \frac{1}{720 \pi^2 }\frac{H^4 k^5}{a^8}  \left(1 + \frac{3}{7} \frac{k^2}{(a H)^2} + \frac{1}{35}\frac{k^4}{(aH)^4} \right) + (\mbox{contact terms})\ .
\end{align}
We observe that the two-point function in the free theory scales as $a^{-8}$, as expected from the $a^{-4}$ scaling of the operator.
In the final line, we have dropped all terms that are analytic in $\k$---e.g.~$(\k^2)^n$, with $n$ being a non-negative integer.  If we Fourier transform such terms back to position space, they become contact terms---i.e.~terms proportional to $\delta(\x)$---and therefore do not contribute to correlation functions at separated points.  Notice that all terms proportional to the cutoff $\Lambda$ are contact terms (as they should be for renormalized operators).  

\vskip 4pt
\noindent
{\it One-loop correction.}---Now consider the non-Gaussian correction to the cross-correlation
\begin{align}
\int \d^3 x \, e^{i \k \cdot \x} \, \langle \cO_R(\x) \phi(\0)\hskip 1pt \rangle &= \frac{1}{a^4} \int^{a\Lambda} \frac{ \d^3 q}{(2\pi)^3} \, q^2 |\k-\q\hskip 1pt|^2 \langle \phi_{\q}\hskip 1pt \phi_{\k - \q} \hskip 1pt \phi_{-\k} \rangle' \ ,
\end{align}
where
\beq
 \langle \phi_{\q}\hskip 1pt \phi_{\k - \q} \hskip 1pt \phi_{-\k} \rangle' =  i \int^\tau_{-\infty} \d \tilde \tau \, a(\tilde \tau)\,   \langle  H_{\rm int}(\tilde \tau)  \phi_{\q} \hskip 1pt \phi_{\k - \q} \hskip 1pt \phi_{-\k}(\tau)  \rangle' + h.c.
\eeq
The notation $\langle \cdots \rangle'$ denotes that an overall delta function has been omitted.
Substituting the interaction Hamiltonian,
\beq
H_{\rm int} =  \frac{1}{ 3 M^2}\int \d^3 x\,  (\phi')^3\ , \label{equ:Hint}
\eeq 
we get
\begin{align}
 \langle \phi_{\q}\hskip 1pt \phi_{\k - \q} \hskip 1pt \phi_{-\k} \rangle' &=  \frac{2 }{M^2}\, \phi_{\q}^*(\tau) \hskip 1pt \phi_{\k - \q}^*(\tau) \hskip 1pt \phi_{-\k}^*(\tau) \, i \int^\tau_{-\infty} \frac{\d \tilde \tau}{H \tilde \tau}   \, \phi_{\q}'(\tilde \tau) \hskip 1pt \phi_{\k - \q}'(\tilde \tau) \hskip 1pt \phi_{-\k}'(\tilde \tau)  + h.c. \ ,
 \end{align}
 where primes stand for derivatives with respect to conformal time.
Since we are interested in the behavior as $q \to \infty$, we keep only the leading terms in $k$, 
\begin{align}
 \langle \phi_{\q}\hskip 1pt \phi_{\k - \q} \hskip 1pt \phi_{-\k} \rangle'  &= \frac{1}{4}\frac{H^5}{M^2} \frac{(1-iq\tau) (1-i|\k-\q\hskip 1pt|\tau) (1-ik\tau)}{q |\k-\q\hskip 1pt| k}  \, i \int_{-\infty}^\tau \d \tilde \tau \, \tilde \tau^2 e^{-iK(\tilde \tau-\tau)} + h.c. \ , \nonumber
 \\
 &= \frac{1}{8}\frac{H^5}{M^2} \frac{1}{k} \frac{1}{q^5} \Big(1+ q^2 \tau^2 + 2 q^4 \tau^4 \Big) \Big(1 + \cO\Big(\frac{k}{q}\Big) \Big)\ , \label{equ:limitX} 
 \end{align}
 where $K \equiv q + |\k-\q| + k$. 
Note the importance of the Bunch-Davies vacuum in deriving eq.~(\ref{equ:limitX}). In an excited state negative frequency modes would lead to contributions with $K \to q - |\k-\q| +k \sim k$.  This would lead to extra inverse powers of $k$. 
Hence, we find
\begin{align}
\int \d^3 x \, e^{i \k \cdot \x} \, \langle \cO_R(\x) \phi(\0)\hskip 1pt \rangle &= \frac{1}{8} \frac{1}{a^4} \, \frac{H^5}{M^2}\, \frac{1}{k}  \int^{a\Lambda} \frac{\d^3 q}{(2\pi)^3} \frac{1}{q} \Big(1+ q^2 \tau^2 + 2 q^4 \tau^4 \Big)\ , \nonumber \\
 &= \frac{1}{32 \pi^2} \frac{1}{ a^2  }\frac{1}{k} \left( \frac{\Lambda^2 H^5}{M^2} +\frac{1}{2}\frac{\Lambda^4 H^3}{M^2} +\frac{1}{3}\frac{\Lambda^6 H}{M^2} \right) + \cO(k^0) \ . \label{equ:cross}
\end{align}
Notice that this cross-correlation scales as $a^{-2}$ and not as $a^{-4}$ (as we would naively expect from the scaling of the operator). The UV divergence has significantly affected the time dependence of the correlation function. This significant change in the scaling behavior of the operator would be a real problem if it weren't possible to remove the contribution by a local counterterm. 

\vskip 4pt
\noindent
{\it Renormalization.}---It is easy to see that the contribution in (\ref{equ:cross}) can be removed by the following local operator 
\beq
\delta \cO \equiv  -  \frac{1}{16 \pi^2} \left( \frac{\Lambda^2 H^5}{M^2} +\frac{1}{2}\frac{\Lambda^4 H^3}{M^2} +\frac{1}{3}\frac{\Lambda^6 H}{M^2} \right)  \frac{\partial^2 \phi}{a^2} \ ,
\eeq  
since
\beq
\int \d^3 x \, e^{i \k \cdot \x} \, \langle \partial^2 \phi(\x) \phi(\0)\hskip 1pt \rangle = - k^2 |\phi_\k|^2 = - \frac{H^2}{2 k} + \cO\left( \frac{k^2}{(aH)^2} \right)\ .
\eeq
Moreover, we can also cancel higher powers of $k$ in the expansion in (\ref{equ:cross}).  The first correction, of order $k^0$, is a pure contact term and therefore doesn't have to be removed explicitly.  In fact, every even power, $(k^2)^n$, where $n$ is a non-negative integer, is a contact term and thus none of these terms contribute to correlation functions at separated points.  This leaves the odd powers, $k^{2m-3}$, where $m$ is a positive integer.  It should be clear that all these terms can be removed by local operators of the form $(\partial^2)^m \phi$.  As a result, the contributions to the correlation function that lead to a physical scaling are associated with $q \sim k \ll a H$ (which we did not compute here).  Clearly, all such contributions are suppressed by $a^{-4}$, as desired.  
In Appendix~\ref{sec:Renormalization}, we will argue that defining renormalized operators by adding local counterterms is always possible in the Bunch-Davies vacuum. These operators have well-defined scaling behavior and are therefore suppressed at late times.
In particular, operators with $n$ derivatives vanish like $(k/a)^n$.
In Appendix~\ref{sec:Renormalization}, we will also show that the renormalized operators satisfy the same symmetries as the bare operators.
Higher-derivative composite operators therefore only make subleading contributions in eq.~(\ref{equ:taylorcurvature}). This completes our proof.


\subsection{Summary of Results}
\label{sec:results}

We have shown that
\beq\label{equ:result}
\lim_{a \to \infty} {\dot {\hat \zeta}}_\k = 0 + \cO\left(\frac{k^2}{a^2}\right) \ .
\eeq
Since this is an operator statement, it applies at all orders in the loop expansion.
This means that any correlation function of $\dot {\hat \zeta}_\k$ will vanish as $a \to \infty$. Equivalently, correlation functions of $\hat \zeta_\k$ are time independent outside the horizon at all-loop order.

\vskip 4pt
We made four important assumptions in establishing this result:  
\begin{enumerate}
\item We assumed that we can transform to a gauge in which the scalar component of the metric, $\zeta$, is the only propagating degree of freedom (in addition to gravitons).  \item We assumed  that the theory is local, in the sense that any pair of local operators $\hat \cO_1$ and $\hat \cO_2$ satisfies 
\beq
\big[\hat \cO_1(\x,t), \hat \cO_2(\y,t) \big] = 0\ , \qquad {\rm for} \quad \x \neq \y\ .
\eeq
\item We assumed that couplings in the action for $\zeta$ depends only logarithmically on the scale factor, i.e. $\lambda(t) \propto ( \log a(t) )^r$, for some non-negative $r$.  
\item We assumed the Bunch-Davies vacuum state.
\end{enumerate}
Using assumptions 1 -- 3, we derived the following operator equation
\beq
\dot {\hat \zeta} (\x,t) = f[\hat \zeta(\x,t)] + \cO(a^{-3}) \ .
\eeq
Invariance under diffeomorphisms required that the lowest order terms in the derivative expansion are given by
\beq
\dot {\hat \zeta} (\x,t) =   \beta_1(t)  \hskip 1pt a^{-2} e^{-2 \hat \zeta}\big( \hskip1pt \partial^2 \hat \zeta + \tfrac{1}{2}  \delta^{ij} \partial_i \hat \zeta \partial_j \hat \zeta \hskip 1pt \big) + \cO(\partial^4) \ .
\eeq
This ensures that every operator in this series is suppressed by at least two derivatives.  
Finally, we showed that if assumption 4 holds, all operators containing derivatives are suppressed by factors of $k/a$ and hence eq.~(\ref{equ:result}) follows. 
\vskip 6pt
Let us ask where our proof would fail if any of these assumptions were violated:
\begin{enumerate}
\item In multi-field inflation, additional light scalars $\sigma$ are present in $\zeta$-gauge.  The proof that $\dot \zeta$ must satisfy an operator equation would still hold, but nothing would forbid terms of the form
\beq
\dot {\hat \zeta}(\x,t)  = \kappa_1(t) \hat \sigma(\x,t)  + \kappa_2(t) \hat \sigma^2(\x,t) + \cdots \ .
\eeq
The fluctuations of $\sigma$ can be non-zero outside the horizon and therefore $\dot {\hat \zeta}_{\k\to 0}$ need not vanish.
More dramatically, ref.~\cite{Endlich:2012pz} recently suggested an inflationary model (solid inflation)  in which the adiabatic mode is completely absent and it isn't possible to go to the standard $\zeta$-gauge.
Our proof then doesn't apply.  In fact, in solid inflation $\zeta$ isn't conserved (even at tree level).

\item Ref.~\cite{Creminelli:2012xb} introduced an inflationary model (Khronon inflation) in which $\zeta$ evolves as 
\beq
\zeta_\k(\tau) \propto \frac{1}{\sqrt{2 k^3}} e^{i \alpha k \tau}\ ,
\eeq
where $\alpha$ is a ratio scales that will not matter here.  We see that $\dot \zeta_{\k\to0} = i (k/a)  \zeta_{\k\to0}$, which violates our eq.~(\ref{equ:result}).  However, one also finds that
\beq
\big[\dot {\hat \zeta}_\k(\tau), \hat \zeta_{\k'}(\tau)\big] = \frac{i}{k^2} \delta(\k + \k') \ ,
\eeq
which is non-local in real space. Khronon inflation therefore violates our locality assumption.

\item 
In the model of ref.~\cite{Namjoo:2012aa}, the coefficient of the kinetic term $\dot \zeta^2$ scales as $a^{-6}(t)$, violating our assumption that couplings in the Lagrangian scale at most as $\log a(t)$.  The authors of \cite{Namjoo:2012aa} then find solutions that scale as $\zeta \propto a^{3}$. This growing mode becomes classical and clearly satisfies an operator equation of the form $\dot {\hat \zeta} \sim 3 H \hat \zeta$.  Why is this equation not forbidden by the dilatation symmetry?  First, we note that, due to the significant time dependence, the commutator scales as $[\dot {\hat \zeta}_\k(t), \hat \zeta_{\k'}(t)] \sim a^3 $ for $a \to \infty$. In this case, our operator equation takes the form $\dot {\hat \zeta} \sim 3 H \hat \zeta + \cO(a^0)$.  This is consistent with the dilatation symmetry because  $\zeta \mapsto \zeta +\lambda$ can be absorbed into $\cO(a^0)$.

\item Assuming the Bunch-Davies vacuum was only important for the renormalization of operators.  In Appendix~\ref{sec:Renormalization}, we show that large corrections to the scaling behavior of composite operators can be removed by a redefinition of the local operator.  This renormalization procedure essentially requires that the only divergences in correlation functions arise from operators at coincident points.  In some excited states, this is known not to be the case~\cite{Banks:2002nv}.  This is usually taken as a sign that the interacting theory is ill-defined.
\end{enumerate}

\section{Relation to the Single-Field Consistency Relation}
\label{sec:Applications}

In the previous section, we used both locality and symmetry to demonstrate that $\dot \zeta$ vanishes outside the horizon as $a^{-2}$.  In the process, we understood the late-time scaling behavior of many other operators. 
 In this section, we will see how this information is useful for understanding the behavior of correlation functions of $\zeta_{\k}$ when there are large hierarchies between the momenta (i.e.~for soft limits).  Specifically, when the operator product expansion (OPE) applies, these correlation functions are determined in terms of the operators with the lowest scaling dimensions.\footnote{The utility of OPEs to describe soft limits of inflationary correlation functions has recently been emphasized by Kehagias and Riotto~\cite{Kehagias:2012pd}.}


\subsection{Operator Product Expansion}


The OPE is a powerful tool for understanding quantum field theories in situations where the scaling behavior of operators is well understood~\cite{weinberg2005theV2}.  The basic idea of the OPE is to replace a set of operators in the neighborhood of a point $\x $ by a sum over local operators at $\x $.  When the operators can be organized according to their scaling dimensions (i.e.~if one knows, for each operator, how many powers of the distance appear in correlation functions), then the leading contribution can be determined by the first few terms in the expansion.  In the case of conformal field theories, one even understands the scaling behavior well enough to re-sum parts of the expansion.
In the context of inflation, we would like to apply the OPE to correlation functions of~$\zeta$.  On the surface, this doesn't look like a well-controlled procedure since $\zeta(\x) \zeta(\0) \sim \log(|\x|)$ and therefore higher powers of $\zeta$ are not suppressed in the OPE.  However, in practice, the OPE is controlled by the smallness of $\langle \zeta^2 \rangle$. 
Furthermore, from the results of the previous sections (and Appendix~\ref{sec:Renormalization}), we will be able to constrain the coefficient functions and/or the scaling behavior of each local operator.  

\vskip 4pt
Consider the following OPE
\beq\label{equ:opedef}
 \zeta(\x) \zeta(\y) \ \xrightarrow{\x \to \y}\ \sum_{\cO} f_{\cO}(x_-)\hskip 1pt \cO(\x_{+}) \ ,
\eeq
where we defined
\beq
\x_{+} \equiv\frac{1}{2}(\x+\y)\quad {\rm and}\quad x_-\equiv |\x-\y|\ .
\eeq
In Fourier space, this OPE reads
\begin{align}
\zeta_{\k- \frac{1}{2}\q} \hskip 1pt \zeta_{-\k-\frac{1}{2}\q} &\ \xrightarrow{|\k| \gg |\q|} \ \sum_{\cO} f_\cO(k)\hskip 1pt  \cO_{-\q} \ . \label{equ:OPE-mom}
\end{align}
The types of local operators $\cO$ that should be included on the right-hand side depend on the field content of theory.  Restricting to single-field inflation, the operators are composite operators made out of $\zeta$ and its derivatives. 
The coefficient functions $f_{\cO}(u)$ (or their Fourier transforms~$f_\cO(k)$) are constrained by the symmetries of Section~\ref{sec:Symmetries}.
Our arguments in Section~\ref{sec:Proof} restrict the appearance of the operator $\dot \zeta$ in the OPE (in particular, up to corrections that vanish as $a^{-3}$, we can replace $\dot \zeta$ using eq.~(\ref{equ:taylorcurvature})).
Moreover, from the discussion in Appendix~\ref{sec:Renormalization}, we know that higher-derivative composite operators are also suppressed by powers of $a(t)$.
The dominant operators in the OPE are therefore operators without derivatives
\beq
 \zeta(\x) \zeta(\y)\ \xrightarrow{\x \to \y}\ \sum_{n} f_{n}(x_-) \hskip 1pt \zeta^n_R(\x_{+}) + \cdots  \ . \label{equ:OPE}
\eeq
Acting $n$ times with the dilatation charge, $[Q_d, \cdots]$, on both sides of eq.~(\ref{equ:OPE}), we find
\beq
f_n(x_-) =\frac{1}{n!}\left( \frac{d}{d\ln x_-}\right)^n \xi(x_-) + \delta_{n 2} \label{equ:fn(x)} \qquad {\rm and} \qquad f_n(k) = \frac{1}{n!}\left( 3+\frac{d}{d\ln k}\right)^n P(k) \ ,
\eeq
where $\xi(x_-) \equiv \langle \zeta(\x) \zeta(\y) \rangle$ and $P(k) \equiv \langle \zeta_\k \zeta_{-\k} \rangle'$.
The near-Gaussianity of the fluctuations, $f_{\mathsmaller{\rm NL}} \zeta \ll 1$, implies that it is often sufficient to keep only the $n=1$ term in the sum
\begin{align}
\zeta_{\k- \frac{1}{2}\q} \zeta_{-\k-\frac{1}{2}\q} &\ \xrightarrow{|\k| \gg |\q|} \  f_1(k)\hskip 1pt \zeta_{-\q } + \cdots\ = \ \frac{d \ln(k^3 P(k))}{d \ln k} \hskip 1pt \zeta_{-\q} + \cdots \ . \end{align}

To capture subleading corrections, we should also consider derivative operators that have finite correlation functions in the limit $a \to \infty$, such as 
\beq
 \zeta(\x) \zeta(\y) \ \xrightarrow{\x \to \y}\ \cdots +   g_1(x_-) \hskip 1pt \x_-^i \hskip 2pt [\partial_{i}\hskip 1pt \zeta](\x_{+}) + g_2 (x_-)\hskip 1pt x_-^2 \hskip 2pt [\partial^2 \zeta](\x_{+}) + \cdots  \ , \label{equ:2point}
\eeq
where the functions $g_1(x_-)$ and $g_2(x_-)$ are at most logarithmic in $x_-$.
The Fourier transform of eq.~(\ref{equ:2point}) is
\beq
\zeta_{\k- \frac{1}{2}\q} \zeta_{-\k- \frac{1}{2}\q} \ \xrightarrow{|\k| \gg |\q|} \  \cdots + \left( g_1(k)\hskip 1pt \frac{\q\cdot\k}{k^2} +  g_2(k) \hskip1pt \frac{q^2}{k^2} \right) \zeta_{-\q} + \cdots   \ , \label{equ:q2}
\eeq
where the functions $g_1(k)$ and $g_2(k)$ scale as $k^{-3}$.  
 Acting with the charge of SCTs, $[Q^i_{sc},\cdots]$, on both sides of eq.~(\ref{equ:q2}) implies that $g_1(k) =0$, which ensures that first subleading term is suppressed by $q^2/k^2$~\cite{Creminelli:2012ed}.

\subsection{Single-Field Consistency Relation}

One is often interested in the coincident limit of correlation functions where two or more operators are brought close to each other (relative to the distances to other operator insertions). The OPE is a natural way to analyze this.  For instance, let us consider the following limit of the three-point function
$\lim_{\x \to \y} \langle \zeta(\0) \zeta(\x) \zeta(\y) \rangle$.  
In Fourier space, this corresponds to the squeezed limit of the bispectrum, 
\beq
\lim_{|\q| \ll |\k|} \big\langle \zeta_{\q}\hskip 1pt \zeta_{\k- \frac{1}{2}\q} \zeta_{-\k- \frac{1}{2}\q} \big\rangle' \ . \label{equ:squeezed}
\eeq  
Maldacena showed that in single-field inflation this limit is fixed by the scale-dependence of the two-point function~\cite{Maldacena:2002vr}. Here, we want to reproduce this result from the OPE. In fact, there is not much left to do.
We simply use the OPE~(\ref{equ:OPE-mom}) to replace the product of the high-momentum modes in (\ref{equ:squeezed}),
\beq
\big\langle \zeta_{\q}\hskip 1pt \zeta_{\k-\frac{1}{2}\q} \hskip 1pt \zeta_{-\k- \frac{1}{2}\q} \big\rangle'  \ \xrightarrow{|\k| \gg |\q|} \ \sum_{\cO} f_\cO(k) \big \langle \zeta_{\q} \hskip 1pt \cO_{-\q} \big\rangle' \ =\ \left(f_1(k)  \ +\cO\left(\frac{q^2}{k^2}\right)  \right) \big\langle\zeta_{\q} \hskip 1pt \zeta_{-\q} \big\rangle'  \ +\ \cdots\  . \label{equ:SqOPE}
\eeq
Terms that aren't shown explicitly in (\ref{equ:SqOPE}) are either suppressed by powers of $a(t)$ or by the near-Gaussianity of the fluctuations.
Hence, we find that the linear operators in the OPE ($n=1$) lead to Maldacena's {\it single-field consistency relation}~\cite{Maldacena:2002vr, Creminelli:2004yq}
\begin{align}
\big\langle \zeta_{\q}\hskip 1pt \zeta_{\k-\frac{1}{2}\q} \hskip 1pt \zeta_{-\k-\frac{1}{2}\q} \big\rangle' \ \xrightarrow{|\k| \gg |\q|} &\   P(q) \left[ \frac{d \ln (k^3 P(k))}{d \ln k} +\cO\left(\frac{q^2}{k^2}\right)\right] P(k) \ .
\label{equ:consistency}
\end{align}
The vanishing of $g_1(k)$ in (\ref{equ:q2}) captures\footnote{For higher $n$-point functions, the conformal consistency relation allows for terms that are linear in the soft external momenta~\cite{Creminelli:2012ed}.  This would arise from an OPE involving $(n-1)$ insertions of $\zeta$ at separated points, but we will not consider such cases here.} the {\it conformal consistency relation} of Creminelli, Nore\~na and Simonovi\'c~\cite{Creminelli:2012ed}, which ensures there is no $\cO(\q \cdot \k)$ correction to (\ref{equ:consistency}).
The subleading corrections (suppressed by $q^2/k^2 \ll 1$) come from operators like the last term in eq.~(\ref{equ:q2}).

In \cite{Assassi:2012zq}, we related the single-field consistency relation to the Ward identity associated with the dilatation symmetry, eq.~(\ref{equ:Ward0}).
Let us make a side remark addressed at readers familiar with that previous work.
There we had to assume that `multi-particle' states (i.e.~states created by acting with several $\zeta$-operators on the vacuum) make a negligible contribution in single-field inflation.  Here, we see that this assumption is equivalent to being able to truncate
the OPE (\ref{equ:OPE-mom}) at order~$n=1$.

\subsection{Violations of the Consistency Relation}

The conservation of $\zeta$ and the consistency relation of the three-point function are closely related. In fact, both are consequences of the non-linearly realized dilatation symmetry. In \S\ref{sec:results}, we list four critical assumptions on which our proof for the conservation of $\zeta$ was based. It is interesting to see how violations of these assumptions map to proposed violations of the consistency relation:

\begin{enumerate}
\item  Of course, it is well-known that large squeezed limits are possible if $\zeta$ isn't the only fluctuating degree of freedom (see e.g.~\cite{Wands:2007bd, Langlois:2012sz} for reviews of non-Gaussianity in multi-field inflation).  In terms of the OPE, these additional fields may appear unsuppressed in (\ref{equ:opedef}).  The coefficients of any such operator is not restricted by the non-linear symmetry of $\zeta$ and therefore can be large.  This is related, but not equivalent, to the violation of the conservation of $\zeta$ we discussed in \S\ref{sec:results}.

\item  In Khronon inflation~\cite{Creminelli:2012xb} the consistency relation is still satisified, but subleading terms are now less suppressed (by $q/k$ rather than $q^2/k^2$). In fact, one could imagine variants of Khronon inflation that would even violate the consistency relation.  It would be interesting to establish a more direct relation between non-locality and the scaling of the bispectrum in the squeezed limit. 

\item Strongly time-dependent couplings (e.g.~power law in $a(t)$ rather than logarithmic) may violate the consistency relation directly through the non-conservation of $\zeta$ \cite{Namjoo:2012aa}.  In this case the operator $\dot \zeta$ scales as $a^n$ (for $n>0$) outside the horizon and is therefore not suppressed  in the OPE.  This is also clear from the argument of Maldacena~\cite{Maldacena:2002vr}, which assumes that the $k = 0$ mode can be removed by a rescaling of the coordinates.  This is only true of time-independent solutions and therefore does not apply in this case.

\item Some excited states have been found to violate\footnote{Technically speaking, the authors of~\cite{ Agullo:2010ws, Agullo:2011xv, Ganc:2011dy, Chialva:2011hc} do not claim to violate the consistency relation in the limit $k\to0$.  However, if we removed the requirement that the states have finite energy, then a true violation can arise.  Such states are essentially the same as those discussed in \cite{Banks:2002nv}. } the consistency relation~\cite{ Agullo:2010ws, Agullo:2011xv, Ganc:2011dy,Chialva:2011hc}.  This can be understood as arising at times before the long mode has crossed the horizon, and hence derivative operators are not suppressed.  However, since we directly apply the OPE at late times, it may not be clear where our argument breaks down.  The resolution lies in the fact that the suppression of composite operators at late times ($a \to \infty$) requires that we can remove any divergent result by a local counterterm.  In these excited states, this is not the case \cite{Banks:2002nv}.  However, this is an unphysical feature of these states that arises because they have infinite energy.  One recovers the consistency relation in the $k \to 0$ limit when restricting to finite energy states~\cite{ Agullo:2010ws, Agullo:2011xv, Ganc:2011dy,Chialva:2011hc}.  For an extensive discussion of these (and related) examples and their observational consequences, see \cite{FGR}.
\end{enumerate}

\section{Discussion}
\label{sec:Conclusions}

In this paper, we proved that the superhorizon conservation of the curvature perturbation $\zeta$ in single-clock inflation holds as an operator statement.  In the process, we developed techniques for understanding correlation functions of $\zeta$ that did not require explicit use of perturbation theory.  We used these insights to understand the operator product expansion of $\zeta$ and its relation to the single-field consistency relation.

There is reason to believe that the technical developments that we used to understand the conservation of $\zeta$ may have applications to other problems.  For example, we have not addressed the conservation of tensor modes.  It should be clear that all constraints that followed from locality alone should apply equally to tensor perturbations.  On the other hand, the transformation properties under large diffeomorphisms are quite different \cite{Hinterbichler:2012nm} and could lead to interesting results.  
Another application would be to eternal inflation.  In this case, the fluctuations of $\zeta$ are order one, which presents a challenge for using traditional perturbation theory techniques.  However, our only result that made explicit use of perturbation theory was the scaling behavior of renormalized composite operators.  For this reason, it is possible that some of our results will survive in the regime of eternal inflation.

Finally, our primary concern was one type of infrared divergence of inflationary correlators, namely those that scale as $\log a(t)$.  There are also infrared divergences that scale as $\log  L$, where $L$ is a hard infrared cutoff on the comoving momenta.  These types of divergences have been studied by many authors (see e.g.~\cite{Seery:2010kh, Giddings:2010nc, Byrnes:2010yc, Gerstenlauer:2011ti, Giddings:2011zd, Giddings:2011ze, Miao:2012xc, Senatore:2012nq}) and it would be interesting to see if our understanding of the conservation of $\zeta$ can shed any additional light on this other class of divergences.

\newpage
\vskip 10pt
\noindent
{\it Note added:} When this paper was completed, ref.~\cite{Senatore:2012ya} appeared which also presents an all-orders proof for the conservation of $\zeta$ on superhorizon scales.

\subsubsection*{Acknowledgements}

We thank Raphael Flauger, Daniel Harlow, Rafael Porto, Leonardo Senatore, Douglas Stanford and Matias Zaldarriaga for helpful discussions.
D.B.~and V.A.~gratefully acknowledge support from a Starting Grant of the European Research Council (ERC STG grant 279617).  The research of D.G.~is supported in part by the Stanford ITP and by the U.S. Department of Energy contract to SLAC no. DE-AC02-76SF00515.  D.G.~thanks the Canadian Institute for Theoretical Astrophysics for hospitality and the opportunity to present this work.

\newpage
\appendix
\section{Renormalization of Composite Operators}
\label{sec:Renormalization}

In quantum field theory in flat space, the Callan-Symanzik equation provides the connection between UV divergences and the scaling behavior of composite operators~${\cal O}(\x,t)$~\cite{collins1984renormalization, weinberg2005theV2}.  In perturbation theory, the scaling of  composite operators is only corrected logarithmically, i.e.~by small anomalous dimensions. This is manifest in renormalization schemes without power law divergences such as dimensional regularization.
In schemes that allow for power law divergences, any power law corrections to the scaling can be removed by introducing {\it local} counterterms in the definition of renormalized operators ${\cal O}_R$.  In this sense, only $\log$ corrections to scalings are physically meaningful.  

In this paper, we have been interested in the behavior of cosmological correlations functions in the limit $a(t) \to \infty$.  Here, the scale factor $a(t)$ plays the role of an infrared regulator and the scaling with time $t$ is controlled by the Hamiltonian and not the renormalization group. 
A priori, it is not obvious that there should be a relation between the results in flat space and in de Sitter space (although the two are mapped to each in the dS/CFT duality~\cite{Strominger:2001pn}).  In this appendix, we will show explicitly that the intuition regarding anomalous dimensions in ordinary field theory will continue to hold for the scaling of cosmological correlation functions.  
 In particular, we will show that the scaling behavior of composite operators is corrected in perturbation theory, at most, by $\log a(t)$.  For example, suppose that,  in the Gaussian theory, the two-point function of some composite operator $\cO(\x,t)$ scales like $a^{-n}$, where $n$ is some integer.  We will show that at higher orders in perturbation theory, any contributions that scales like $a^{-m}$, where $m$ is an integer with $m < n$, 
can be removed by a local redefinition of the operator, 
\beq
\cO_R(\x,t) \equiv \cO(\x,t) + \delta \cO(\x,t)\ .
\eeq 


\subsection{Renormalizability by Local Counterterms}
\label{sec:Counter}

In Section~\ref{sec:Proof}, we presented a specific example for the renormalization of a composite operator by local counterterms.
This example was meant to be illustrative, but it does not establish that this procedure works at all orders in perturbation theory.
In this section, we will put forward arguments to that effect.

\subsubsection{Momentum Space Argument}

Perturbation theory is formulated most straightforwardly in momentum space, while locality is most manifest in position space.
We will therefore present the argument twice, here in momentum and below in position space. In each case, we will first present the general strategy and then the details of the `proof'.

\vskip 4pt
\noindent
{\it Strategy.}---Consider the composite operators\footnote{For notational simplicity, we will write most expressions for a specific composite operator, but our results hold for any local operator constructed from $\phi$ and derivatives. } $\cO(\x,t) = \prod_{i=1}^n (a^{-2} \partial^2)^{r_i} \phi(\x,t)$. In momentum space, this becomes
\beq\label{equ:localmomentum}
\cO_{\k}(t) \ \equiv\  \prod_{i=1}^n \int \frac{\d^3 k_i}{(2\pi)^3} \left( \frac{k_i^2}{a^2} \right)^{r_i} \phi_{\k_i}(t) \, \delta\Big(\sum_i \k_i - \k\Big)\ .
\eeq 
Regions of finite momentum in the integrand of (\ref{equ:localmomentum})
scale  like $a^{- \sum_i 2 r_i}$ in the limit $a \to \infty$, relative to the correlation function of $\prod_{i=1}^n \phi_{\k_i}$.  Therefore, any contributions that do not scale like $a^{- \sum_i 2 r_i}$, must come from momentum configurations with $k_i \to \infty$ and $k_i/a$ fixed (assuming that $\sum r_i >0$).
Because momentum is conserved, a least two of these momenta must diverge together.  For purpose of illustration, let us consider the case where all the $k_i$'s diverge as $a\to \infty$.  
We are interested in the scaling behavior of $\cO_\k$ inside of correlation functions, such as
\beq
\lim_{a \to \infty} \big \langle \cO_\k \,  \phi_{\p_1} \cdots \phi_{\p_m} \big\rangle  =\lim_{a \to \infty} \Big\langle \, \prod_{i= 1}^n \int \frac{\d^3 k_i}{(2\pi)^3} \Big(\frac{k_i^2}{a^2} \Big)^{r_i} \phi_{\k_i} \delta\Big(\sum_i \k_i - \k \Big) \,  \phi_{\p_1} \cdots \phi_{\p_m} \Big\rangle \ . \label{equ:integral}
\eeq
Our goal is to show that
\beq
\lim_{a, k_i \to \infty} \Big\langle \, \prod_{i= 1}^n \Big(\frac{k_i^2}{a^2} \Big)^{r_i} \phi_{\k_i} \delta\Big(\sum_i \k_i - \k \Big) \,  \phi_{\p_1} \cdots \phi_{\p_m} \Big\rangle \ = \ F(k_i, a) \times \prod_{j=1}^m p_j^{-3} \left[ 1+ \cO\Big(\frac{p^2}{a^2}\Big) \right] \ . \label{equ:goal}
\eeq
Let us explain why this is the desired result.  
First, notice that (\ref{equ:goal}) factorizes into a function of the diverging momenta $k_i$ and a function of the finite momenta $p_j$.  This means that the integrals over the momenta $k_i$ in (\ref{equ:integral}) will simply give a number ${\cal C}$ times the function of the momenta $p_j$.  Second, the remaining function of $p_j$ is itself a correlation function of the fields $\phi_{\p_j}$ and some local operator at some lower order in perturbation theory.  In the example above, the leading term would be the correlation function of $\int \d^3 x\, e^{i \k \cdot \x} \phi^m(\x)$ and $\phi_{\p_1} \cdots \phi_{\p_m}$. 
On the other hand, if we were to find a factorized answer that contained higher inverse powers of $p_j$ (e.g. $p_j^{-5}$), then we would not be able to remove it by subtracting a local operator.  Instead, we would  need to subtract a non-local operator containing powers of $\partial^{-2}$.  

Our strategy will be to show that each Feynman diagram contributing to the above correlation function can be factorized into a sub-diagram containing the diverging momenta and one containing only finite momenta.  After integrating over the momenta $k_i$, we can simply replace the divergent sub-diagram with a local operator.  

In the following, we will generalize the standard arguments from flat space quantum field theory (see Weinberg, Vol.~II, Ch.~20~\cite{weinberg2005theV2}) to de Sitter space and the associated $in$-$in$ correlation functions.
Our argument will fall short of being a complete proof for technical reasons related to the regions of integration of the loop momenta.  This is the same complication that arises in the standard arguments for validity of the OPE in flat space, like those in~\cite{weinberg2005theV2}.  We will explain this in more detail at the end of this subsection.

\vskip 6pt
\noindent
{\it `Proof'.}---For concreteness, let us consider an $(n+m)$-point function of the form
\beq
\big\langle \left(a^{-2}\partial^2 \phi\right)_{\k_1} \cdots \left( a^{-2}\partial^2 \phi \right)_{\k_n} \phi_{\p_1} \cdots \phi_{\p_m} \big\rangle\ , \label{equ:17}
\eeq
where all the operators are evaluated at some fixed time $t$.  
We are interested in the behavior as $k_i \to \infty$ and $a \to \infty$ with $k_i / a$ fixed. 
One important feature of inflationary correlation functions is that the metric contains a trivial rescaling symmetry $a \to \lambda a$ and $\x\to \lambda^{-1} \x$ (and hence $\k \to \lambda \k$).  Under this symmetry, local scalar operators transform as $\lambda^0$ and therefore their momentum space counterparts transform as $\lambda^{-3}$.  In general, this symmetry is broken in the action for $\zeta$ because any explicit function of $t$ can be rewritten in terms $\log a(t)$.  In other words, we can't rescale $a$ and $t$ independently after solving for the background.  We will assume that the time dependence of the couplings is at most logarithmic in $a(t)$ (power law in $t$, during inflation) such that the power law scaling, $\lambda^{-3}$, is valid up to small corrections.  Usually, this is required in order to preserve the near scale invariance of the fluctuations. As a result, we can think of the momentum dependence\footnote{The mode functions in a scale-invariant theory are of the form $\phi_{\k}(t) \sim k^{-3/2}f(k/a(t))$.  Here, we are associating a $k^{-3/2}$ scaling with the creation and annihilation operators.} of any field as being $k^{-3} f(k/a)(1+ \tfrac{1}{2}(n_s-1) \log k+\cdots)$, where $k/a$ is counted as~$k^0$.  

Let us consider Feynman diagrams associated with the following $in$-$in$ correlation function
\beq\label{equ:ininmomentum}
\int_{-\infty}^t \d t_1  \, \cdots \int_{-\infty}^{t_{r-1}} \d t_r \, \big\langle \big[H_{\rm int}(t_1), \, \ldots , \big[H_{\rm int}(t_r) , \left(a^{-2}\partial^2 \phi\right)_{\k_1} \cdots \left( a^{-2}\partial^2 \phi \right)_{\k_n} \phi_{\p_1} \cdots \phi_{\p_m}\big] \big] \big]  \big\rangle\ .
\eeq
We will focus on contributions from a diagram or sub-diagram, $\Gamma$, in which all internal momenta are of order $k_i$ (for simplicity, we will take all the $k_i$'s to be of comparable magnitude).  
Every external line contributes a factor\footnote{For convenience, we will assume exact scale invariance of the interaction picture fields.} of $k_i^{-3}$ or $p_j^{-3}$.  Since $H_{\rm int} = \int \d^3 x\, a^3(t) {\cal H}_{\rm int}(\x,t)$ in position space, when written in momentum space, there is a momentum integral for every field and an overall momentum-conserving delta function.  In terms of Feynman rules, this means that every internal line contributes a factor of $k_i^{-3}$ for the contraction of the interaction picture fields and an integral $\int \d^3 k_i$ (recall that derivatives of the fields scale as $k/a \sim k^0$).
Since every vertex is associated with an insertion of $H_{\rm int}$, each vertex introduces a momentum-conserving delta function and a time integral $\int \d t' a^3(t')$, where we will count the factor of $a^3$ as $k^3 f(k/a)$.  In addition, there is a commutator associated with each vertex coming from the $in$-$in$ expression~(\ref{equ:ininmomentum}).  
Each commutator with the fields $\phi_{\p_j}$ is suppressed by $p_j^3/a^3$. These contributions can be ignored in the limit $a \to \infty$.
On the other hand, commutators acting on the internal lines do not affect the scaling since we are counting $k/a$ as having scaling $k^0$.  

Consider a general diagram with $N$ vertices, $I$ internal lines, $E_k$ external $k$ lines and $E_p$ external $p$ lines.  The overall momentum scaling of the diagram is $\Gamma \sim k^{D}$, where 
\beq
D = -3 E_k + 3 I - 3 I + 3 N - 3(N-1) = -3 E_k + 3\ .
\eeq
Here, the term proportional to $(N-1)$ comes from extracting the overall momentum-conserving delta function. In writing this expression, we haven't been careful about the scaling with $a(t_\ell)$ where $t_\ell$ ($\ell = 1,\hskip 1pt \cdots \hskip -1pt,r$) is the time appearing the $\ell$-th insertion of $H_{\rm int}$.  We are taking $a(t) \to \infty$, but these integrals run over all values of $t_\ell$, not just the far future.  We have implicitly assumed that the $t_\ell$-integrals receive their dominant contributions at late times, when $a(t_\ell) \to \infty$.
  Fortunately, this assumption is valid in the Bunch-Davies vacuum, because every internal lines comes with a factor of $e^{- \epsilon k/a(t')}$ from the $i\epsilon$ prescription.  As we take $k \to \infty$, only contributions with $a(t_\ell) \gtrsim \epsilon k$ aren't exponentially suppressed, and we may assume $a(t_\ell) \to \infty$ inside $\Gamma$.

Using these Feynman rules, we find that the diagram scales as
\beq
\Gamma \ \sim\ K^3 f\left(\frac{k_i}{a(t)}\right) \left(\log \frac{k_i}{H}\right)^w (\log a(t))^v \times \prod_{i=1}^n k_i^{-3} \prod_{j=1}^m p_j^{-3}  \times \delta \Big(\sum_i \k_i + \sum_j \p_j \Big)\left(1 + {\cal O}\left(\frac{p}{k}\right)\right)\ , \label{equ:Gamma}
\eeq
where $K^3$ is some product of three $k_i$'s and $w,v$ are positive integers. The factors of $\log(k_i/H)$ can arise from the time dependence of the coupling constants.  We do not include $\log(p / H)$ scalings from time-dependent couplings, as they could only arise from early times.  For Bunch-Davies initial conditions, these contributions are exponentially suppressed.  Finally, we have included additional factors of $\log a(t)$ which, in principle,  can arise from the time integration $\int^t \d t' \sim t \simeq H^{-1} \log a$ \cite{Weinberg:2006ac}.

We notice that eq.~(\ref{equ:Gamma}) is precisely of the form of eq.~(\ref{equ:goal}).
We can therefore replace this whole diagram or sub-diagram by a local operator connected to the external lines with momenta $p_j$ times a function $F(k_i)$ (this is the operator production expansion in momentum space). When we define the correlation function in terms of the original local operator, we perform the integral over $\int \d^3 k_1 \cdots \d^3 k_n\, \delta(\sum_i \k_i - \sum_j \p_j)$.  
Because the diagram factorizes, this is simply a number times a local operator.

The higher-order terms in the $p/k$ expansion come from Taylor expanding the internal lines in terms of $p_j$.  For example, an internal line might have momentum $\q = \k_i + \p_j \sim \k_i$.  Using rotation invariance, we can rewrite the dependence on $\q$ in terms of $q \sim (k_i^2 + p_j^2 + 2\k_i \cdot \p_j)^{1/2}$.  If we Taylor expand $q$ in powers of $p_j$, the only odd powers of $p_j$ appear in the combination $\k_i \cdot \p_j$. But odd powers of  $\k_i \cdot \p_j$ vanish when we perform the angular integrals over the $k_i$'s in (\ref{equ:localmomentum}).  With only even powers of $p_j$ surviving in the Taylor expansion, we can remove the entire series by adding derivatives inside the local operator.  Therefore, all contributions to $\Gamma$ that introduce power law changes in the scaling behavior of $\cO$ can be removed by adding local counterterms. 

One may be concerned that the diagram $\Gamma$ could be disconnected, in the sense that not all of the momenta $k_i$ are connected to each other via some path in $\Gamma$.  Let us assume that this is true, i.e.~let us assume that $\Gamma$ is disconnected or connected only through soft internal lines.  For momentum conservation to hold, this would imply that $\sum_{i=1}^{r < n} \k_i \sim \cO(\p_j)$.  This corresponds to a special momentum configuration where some subset of the diverging momenta separately
sum to a finite momentum.  For generic momenta this cannot arise and the diagram must therefore be connected.  Since we will be integrating over $k_i$, these non-generic points can be ignored.  
 
\vskip 6pt
\noindent
{\it Loopholes and caveats.}---The above argument seems very general, so it is worth highlighting situations where it could fail.  First of all, the $i\epsilon$ prescription of the Bunch-Davies vacuum was crucial for suppressing contributions at early times. 
For some excited states this may not be the case.  This is consistent with the observation that some excited states in de Sitter are known not to be renormalizable by local counterterms~\cite{Banks:2002nv}.  In our language, these examples correspond to contributions when $a(t_\ell) \sim p_j / H$, which would induce additional inverse powers of $p_j$ and therefore cannot be removed by local counter-terms.  In extreme situations, one could also imagine compensating for the exponential suppression, $e^{- \epsilon \frac{k}{a H}}$, in the limit $a \to 0$, with exponential growth of the couplings of the form $\lambda(t) \sim \exp(a^{-\delta}(t))$ for $\delta>1$. Having coupling grow this rapidly in the far past would give a large non-local contribution to late-time correlators.  We have excluded such large time dependences throughout the paper.

Finally, we want to stress that our argument falls short of a formal proof, as we have not been careful enough regarding the integration over loop momenta.  Although we have treated all large momenta as order $k_i$, there are always regions of integration where the internal momenta are much larger or much smaller than $k_i$.  One might worry that the result of performing and regulating these integrals might somehow result in additional factors of $k_i / p_j$.  The general scaling behavior of the internal lines makes it difficult to see how such contributions could arise, but we leave a complete investigation to future work.

\subsubsection{Position Space Argument}

In  translationally invariant theories, explicit calculations are often easier to perform in momentum space.  However, locality is a fundamental property of most theories that is easier to understand in position space.  For this reason, we will now explain how the renormalization of composite operators works in position space.  The results will be less explicit than those of the previous section, but may be more intuitive.

\vskip 4pt
\noindent
{\it Strategy.}---Consider the $in$-$in$ master formula
\beq
\langle \cO(\x,t) \cO(\y,t)\rangle = \langle 0 |\, \bar{\rm T}\, e^{ i \int_{-\infty}^t \d t' \,  H_{\rm int}(t')   }  \cO(\x,t) \cO(\y,t)  \, {\rm T} \, e^{- i \int_{-\infty}^t \d t' \,  H_{\rm int}(t')  } \, | 0\rangle \ . \label{equ:master}
\eeq
In perturbation theory this expression is evaluated by expanding in powers of 
\beq
\int \d t' H_{\rm int}(t') =\int \d t' \d^3 x' a^3(t') {\cal H}_{\rm int}(\x\hskip1pt{}',t') \label{equ:INT}
\eeq and using contractions of the interaction picture fields.  We then look at the regions of integration over the positions of ${\cal H}_{\rm int}(\x\hskip1pt{}',t')$ that are not suppressed by powers of $a(t)$.  We will show that these contributions arise from spacetime regions where some number of ${\cal H}_{\rm int}(\x\hskip1pt{}',t')$ are within a sphere surrounding the composite operator $\cO(\x,t)$ that is much smaller than the distances to other operators in the correlation function.  
By Taylor expanding ${\cal H}_{\rm int}(\x\hskip1pt{}',t') = {\cal H}_{\rm int}(\x,t) + (\x\hskip1pt{}'-\x) \cdot \partial_\x \, {\cal H}_{\rm int}(\x,t') + \cdots$, we can then treat ${\cal H}_{\rm int}(\x',t')$ as an operator at the point $(\x,t)$.
To evaluate the behavior at coincident points, such as $\cO(\x,t) \hskip 1pt {\cal H}_{\rm int}(\x, t)$ (and derivatives therefore), we simply use the free field contractions of the fields that make up ${\cal H}_{\rm int}$ and $\cO$.  The result is therefore a new local operator with a divergent coefficient.  The entire contribution can then be removed by adding a local counterterm to $\cO(\x,t)$.  

The argument in position space  is more subtle for the usual reason that perturbation theory is easier to implement in momentum space.  For this reason, let us focus an a sightly simplified correlation function, namely the two-point function of the operator ${\cal O}(\x,t) \equiv (a^{-2}\partial^2 \phi)^n (\x,t)$\hskip 1pt:
\beq\label{equ:ininposition}
\int_{-\infty}^t \d t_1  \, \cdots \int_{-\infty}^{t_{r-1}} \d t_r \, \big\langle \big[H_{\rm int}(t_1), \, \cdots , \big[H_{\rm int}(t_r) , {\cal O}(\x,t)\, {\cal O}(\y,t)\big] \big] \big]  \big\rangle\ .
\eeq
Despite the reduced complexity of the correlation function, it still is sufficient for our main goal.  In particular, if the correlation function between $(a^{-2}\partial^2 \phi)^n (\x,t)$ and any local operators is not suppressed (and cannot be removed by a local counterterm), then we can insert a complete set of states to find that $\langle (a^{-2}\partial^2 \phi)^n (\x,t) | n\rangle$ is unsuppressed for some state $|n\rangle$.  Inserting the same set of states in (\ref{equ:ininposition}) implies that there must be a similarly unsuppressed contribution to this two-point function.  Each such contribution to (\ref{equ:ininposition}) takes the form $|\langle (a^{-2}\partial^2 \phi)^n (\x,t) | n\rangle|^2$ and therefore cannot be cancelled to make the final result vanish.  Hence, it is sufficient to show that (\ref{equ:ininposition}) vanishes in the limit $a \to \infty$ to ensure that correlation functions of ${\cal O}(\x,t)$ with any local operator will be suppressed by powers of $a^{-1}$.  

\vspace{0.5cm}
\begin{figure}[h!]
   \centering
       \includegraphics[scale =0.8]{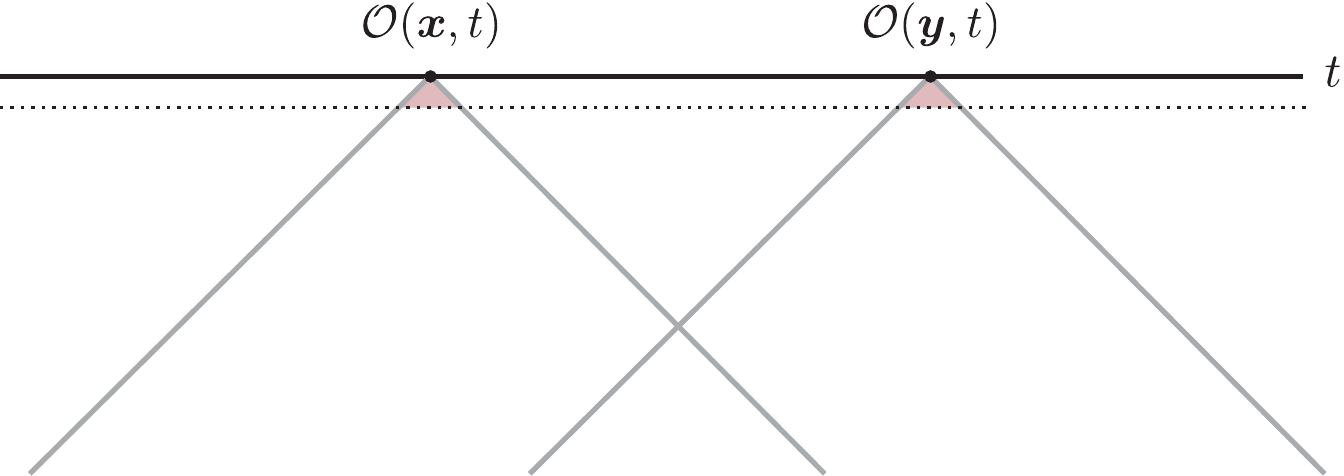}
   \caption{Composite operators $\cO(\x,t)$ and $\cO(\y,t)$ inserted at future infinity of de Sitter space. Locality requires that the operators are only influenced by sources that are inside their past light cones (shown in grey).
The shaded regions near the operators indicate where an insertion of ${\cal H}_{\rm int}(\x',t')$ yields an unsuppressed contribution to $\langle \cO(\x,t) \cO(\y,t) \rangle$. Divergences can therefore be removed by adding local counterterms.}
  \label{fig:RealSpace}
\end{figure}

\vskip 4pt
\noindent
{\it `Proof'.}---Let us first consider the contribution to (\ref{equ:ininposition}) from the insertion of ${\cal H}_{\rm int}(\x_r,t_r)$ that includes a commutator acting on $(\partial^2 \phi)^n$ at $\x$ or $\y$.  Locality requires that this commutator vanish when $(\x_r,t_r)$ is outside the past light-cone of $(\x,t)$ or $(\y,t)$.  When $t_r \sim t$, locality requires that the commutator is proportional to a delta function, $\delta(\x_r - \x) \widetilde \cO(\x,t)$, or derivatives thereof (when acting on $\cO(\x,t)$).  Any such contribution to the correlation function is manifestly local at $\x$ and can be removed by subtracting $\widetilde \cO(\x,t)$.

Potentially dangerous contributions must come form points where $t_r \ll t$.  In the Bunch-Davies vacuum, these contributions are localized on the past light cone of the operator $\cO(\x,t)$, i.e.~$|\x_r -\x| \sim (a(t_r) H)^{-1}$.  This allows us to perform the integral $\int \d^3 x_r \to (a(t_r) H)^{-2} \int \d\Omega$.  The factor of $\int \d t_r\, a^3(t_r)$ from the measure in (\ref{equ:INT}) ensures that these contributions scale as  $a(t_r)$ in the limit $a(t_r) \to 0$. Naively, this suppression by $a(t_r)$ in the measure could be compensated by the divergent contributions along the light-cone arising from the contractions of free fields.  However, the $i\epsilon$ prescription ensures that the only physical divergences come from operators at coincident points.  Specifically, the $i \epsilon$ prescription is equivalent to the analytic continuation of the correlators from the Euclidean sphere.  These correlators are suppressed by powers of the Euclidean distance, which only vanishes at coincident points.  The same will therefore be true of the Lorentzian correlations in the Bunch-Davies vacuum.  Finally, the full expression must be invariant under the rescaling $a \to \lambda a$ and $\x \to \lambda^{-1} \x$.  Due to the power law suppression along the light-cone\footnote{Otherwise, $|\x_r -\x| a(t_r)$ would be consistent with the rescaling symmetry and is unsuppressed.}, our  final result must be suppressed by (at least) $a(t_r) / a(t)$ for $t_r \ll t$.  

So far, we have only considered the contribution to the correlation function from a single insertion ${\cal H}_{\rm int}(\x_r,t_r)$.  We found that, after integrating over $\x_r$ and $t_r$, only points near $(\x,t)$ or $(\y,t)$ contribute significantly (i.e.~do not vanish as $a \to \infty$).  However, we should also consider what happens when the other interactions ${\cal H}_{\rm int}(\x_i,t_i)$ are included.  From the above argument, the contribution from $t_r \sim t$ can still be removed by redefining the local operator, even in the presence of the additional interactions.  This would suggest that the only contributions that cannot be removed are from points with $t_r \ll t$.  However, we found that these points are suppressed by powers of $a$.  The only way we would get a significant contribution would be if the integral over $t_i$ with insertion ${\cal H}_{\rm int}(\x_i,t_i)$ would diverge as a positive power of $a(t)$.  However, for couplings that scale at most as $\log a(t_i)$, the higher orders may diverge at most as $(\log a(t))^r$~\cite{Weinberg:2006ac}.  This completes our demonstration that composite operators can always be renormalized by adding local counterterms.

\subsection{Symmetries of Renormalized Operators}

We conclude this appendix by showing that the basis of renormalized operators transforms under symmetries in the same way as the bare operators.

\vskip 4pt
First, let us review how to analyze the transformation of renormalized operators  in the path integral formalism.  Consider an action $S_0$ that describes the full theory.  To compute correlation functions, we deform the action by
\beq
S \, =\, S_0 + S_n(J) \ ,
\eeq
where
\beq
S_n(J) \equiv -\, \sum_n \int \d^4 x \sqrt{-g} \, J_n(\x,t) \hskip 1pt \zeta^n(\x,t)\ .
\eeq
We have added sources $J_n$ to the action such that correlation functions of $\zeta^n$ can be computed as
\beq
\big \langle \zeta^n(\x,t) \, \cdots \big\rangle_{J_n = 0} =  \frac{\delta}{\delta J_n} \langle \cdots \rangle \Big|_{J_n=0}\ .
\eeq
As far as the action is concerned, $J_n(\x,t)$ is just like any other coupling.  When we perturb in $J_n$ we will therefore find divergences that need to be removed by adding source-dependent counterterms 
\beq
S_R = S_0 + S_n(J) + \delta_c S(J)\ .
\eeq  
Because the theory is now finite, the correlation functions are also finite
\beq
\big\langle [\zeta^n(\x,t)]_{R} \, \cdots \big\rangle_{S_R , J_n = 0} =  \frac{\delta}{\delta J_n} \langle \cdots \rangle \Big|_{S_R, J_n=0}\ . \label{equ:COR}
\eeq
The operators appearing in (\ref{equ:COR}) are therefore the renormalized composite operators $[\zeta^n]_{R} \equiv \zeta_R^n$.
In order to maintain the symmetries of the action, the sources $J_n$ can be given transformations under the symmetries such that $S$ is invariant.  Moreover, we can also choose these transformations to leave $S_n$ invariant.   By taking the functional derivative of $S_n$, we see that the operator $\frac{\delta}{\delta J_n}$ must transform in the same way as $\zeta^n(\x,t)$.  Hence, as long as $ \delta_c S(J)$ does not explicitly break the symmetries, then $S_R$ is also invariant and therefore $\frac{\delta}{\delta J_n} \equiv [\zeta^n]_{R}$ must transform in the same way as $\zeta^n$.

All of this applies equally to cosmological $in$-$in$ correlation functions. In fact,
$in$-$in$ calculations are just a special case of the above analysis, in which the (conformal) time integral goes from $\tau = -\infty(1-i \epsilon)$ to $\tau = -\infty(1+i \epsilon)$, while passing through $\tau = 0$.  Because the symmetries of Section \ref{sec:Symmetries} are continuously connected to the group of diffeomorphisms, the counterterm action $ \delta_c S(J)$ would have to explicitly break diffeomorphism invariance to violate them. 
Using dimensional regularization and the results of the previous subsection, there is no need to use such a regulator.  We therefore conclude that the transformation properties of renormalized operators follow from~(\ref{equ:LG1}) and (\ref{equ:LG2}).

\vskip 6pt
For the specific application in Section \ref{sec:Proof}, there is, in fact, a more direct way to understand the symmetries of renormalized operators.  In \S\ref{sec:symmconstraints}, we used symmetries to forbid the operators that transform non-linearly under the charge $Q_d$.  We then argued in \S\ref{sec:subcomposite} that the remaining operators vanish as $a^{-2}$ after renormalization.  The concern is that renormalization might mix these two groups of operators.  For example, this would arise if $\cO_R(\k) = {\cal L}_{\k} - c\hskip 1pt \zeta_{\k}$, where $\cO_R$ scales as $a^{-2}$, ${\cal L}$ is some operator that transforms linearly under $Q_d$ and $c$ is a constant.  If this were the case, then we would find
\beq
i \big\langle \big[Q_d, \cO_R(\k) \big] \big\rangle = c\hskip 1pt (2\pi)^3\delta(\k) \ .
\eeq
However, from eqs.~(\ref{equ:constantmode1}) and (\ref{equ:constantmode2}), we see that any such mode has a constant contribution to its power spectrum.  This violates the assumption that $\cO_R(\k) \sim a^{-2}$, and therefore we must have $c =0$.  
Repeating this argument for any other operators that vanish as $a\to \infty$, we find that they all transform linearly under the dilatation symmetry.  

\newpage
\addcontentsline{toc}{section}{References}

\bibliographystyle{utphys}
\bibliography{articles,books,website,lecturenotes}

\end{document}